\documentclass[fleqn,usenatbib]{mnras}

\usepackage{newtxtext,newtxmath}

\usepackage[T1]{fontenc}

\DeclareRobustCommand{\VAN}[3]{#2}
\let\VANthebibliography\thebibliography
\def\thebibliography{\DeclareRobustCommand{\VAN}[3]{##3}\VANthebibliography}

\usepackage{graphicx}	
\usepackage{amsmath}	
\usepackage{CJK}
\usepackage{float}
\usepackage{subcaption}
\usepackage{color}

\newcommand{\gd}{GD56}
\newcommand{\wdone}{WD\,1150$-$153}

\newcommand{\Teff}{$T_\mathrm{eff}$}                 
\newcommand{\logg}{$\log(g)$} 
\newcommand{\err}{\,$\pm$\,}
\newcommand{\mum}{$\mu$m~}
\newcommand{\forsterite}{Mg$_2$SiO$_4$} 
\newcommand{\enstatite}{MgSiO$_3$} 

\title[Silicate mineralogy and bulk composition]{Silicate mineralogy and bulk composition of exoplanetary material in polluted white dwarfs}

\author[L. K. Rogers]{Laura K. Rogers$^{1,2}$\thanks{E-mail: laura.rogers@ast.cam.ac.uk}, 
Amy Bonsor,$^{1}$
\'Erika Le Bourdais,$^{3}$ 
Siyi Xu \begin{CJK*}{UTF8}{gbsn}(许\CJKfamily{bsmi}偲\CJKfamily{gbsn}艺)\end{CJK*},$^{2}$
Kate Y. L. Su,$^{4}$
\newauthor Benjamin Richards,$^{1}$
Andrew Buchan,$^{5}$
Nicholas P. Ballering,$^{4}$
Marc Brouwers,$^{1}$
Patrick Dufour,$^{3}$
\newauthor Markus Kissler-Patig,$^{6}$
Carl Melis,$^{7}$
Ben Zuckerman$^{8}$
\\
$^{1}$ Institute of Astronomy, University of Cambridge, Madingley Road, Cambridge CB3 0HA, UK \\
$^{2}$ NOIRLab, 950 N Cherry Ave, Tucson, AZ, 85719, USA \\
$^{3}$ Trottier Institute for Research on Exoplanets and Department of Physics, Universit\'e de Montr\'eal, 1375 Ave. Th\'er\`ese-Lavoie-Roux Montr\'eal, QC \\~~~~H2V 0B3, Canada \\
$^{4}$ Space Science Institute, 4765 Walnut St, Suite B Boulder, CO 80301, USA \\
$^{5}$ Department of Physics, University of Warwick, Coventry CV4 7AL, UK \\
$^{6}$ European Space Agency - European Space Astronomy Centre, Camino Bajo del Castillo, s/n., 28692 Villanueva de la Canada, Madrid, Spain \\
$^{7}$Department of Astronomy \& Astrophysics, University of California San Diego, La Jolla, CA 92093-0424, USA \\
$^{8}$ Department of Physics and Astronomy, University of California, Los Angeles, CA 90095-1562, USA \\
}

\date{Accepted XXX. Received YYY; in original form ZZZ}

\pubyear{2025}

\begin{document}
\label{firstpage}
\pagerange{\pageref{firstpage}--\pageref{lastpage}}
\maketitle

\begin{abstract}

White dwarf planetary systems uniquely link the bulk elemental composition of exoplanetary material to the mineralogy as photospheric abundances can be compared to circumstellar dust mineralogy. This study re-examines \textit{Spitzer}/IRS spectra of eight white dwarfs with both circumstellar dust and photospheric metals. All systems show 10\,\mum silicate emission features consistent with a mixture of olivine and pyroxene silicates, with varying dominance. New \textit{Hubble Space Telescope} ultraviolet spectroscopic observations of two of these systems, \gd{} and \wdone, reveal that both are accreting dry, rocky material. \wdone{} is accreting material consistent with Bulk Earth, while \gd{} is accreting core-rich material with an inferred core mass fraction of 0.59$^{+0.08}_{-0.09}$ (0.37$^{+0.08}_{-0.08}$ by mole). A comparison between the bulk elemental composition of the accreted planetary material and the dust mineralogy of the eight systems reveals a tentative correlation between the dominant silicate mineralogy and the Mg/Si ratio, indicating that the circumstellar and photospheric material are compositionally similar. This suggests that rapid and well-mixed accretion is occurring with minimal compositional alteration. Furthermore, new \textsc{ggchem} equilibrium chemistry models confirm that Mg-rich planetary material preferentially forms olivine-rich dust, highlighting the importance of equilibrium in planetary chemistry and that a host star or rock's Mg/Si can be used to predict whether its silicate mineralogy is olivine- or pyroxene-dominated, influencing its capacity to structurally store water, recycle key nutrients, and possibly habitability.

\end{abstract}




\section{Introduction}

Rocky bodies contain a huge diversity of minerals. On Earth there are thousands of minerals, however, the majority of rocks are formed from a handful of common materials, including the common silicate minerals: olivine, pyroxene, and quartz. These minerals are made from the key rock forming elements: Mg, Si, Fe, and O, which make up more than 90 per cent of Bulk Earth \citep{mcdonough2003compositional}. The mineralogy of a planet's crust and mantle govern its properties. For example, the Earth's mantle is made from a combination of olivine ((Mg,Fe)$_2$SiO$_4$) and pyroxene ((Mg,Fe,Ca)SiO$_{3}$). Mantle mineralogy controls the mantle viscosity and melt fraction, with sweeping consequences for its ability to store water, the rate of volcanic activity, and supply of nutrients to life at the surface, all of which are critical to its habitability \citep[e.g.][]{Kelley2010mantle,Lambart2016role,Wang2018elemental}.

Constraining the composition and mineralogy of exoplanets is an emergent field of study \citep{Foley2024geology}, with several methods being used. One method uses stellar abundances as inputs to formation models to infer the mineralogy of rocky exoplanets \citep{Hinkel2018star,Putirka2019composition,Spaargaren2023plausible} under the assumption that the star and its planet formed from the same refractory composition. This assumption is supported by studies of polluted white dwarfs in wide binaries with main sequence stars \citep{Bonsor2021Host,Aguilera2025Host}. Observations of protoplanetary and debris discs can reveal the composition and mineralogies of the building blocks of exoplanets \citep[e.g.][]{Mittal2015spitzer}. Bulk mass and radii measurements of exoplanets help to constrain their interior structure and composition but have significant degeneracies between models \citep[e.g.][]{seager2007mass, dorn2015can}. Ultra-short period planets orbit so close to their host star that their rocky surfaces evaporate, enabling the study of their surface layers via transmission spectroscopy \citep[e.g.][]{Rappaport2012possible,Bodman2018inferring} with recent \textit{JWST} observations of K2-22b being consistent with a mantle dominated by magnesium silicate minerals \citep{Tusay2025disintegrating}. Lava worlds are less extreme versions of ultra-short period planets and are expected to have atmospheres dominated by vaporised magma allowing for constraints on the mantle composition through studies of their atmospheres \citep[e.g.][]{Schaefer2009chemistry,Ito2015theoretical,Zieba2022k2}. Airless rocky exoplanets can also reveal surface geology from studies of their thermal emission as a function of wavelength \citep{Hu2012theoretical,Paragas2025new}. These methods all make progress towards measuring the composition and mineralogy of exoplanetary material, however, no single method can constrain both elemental composition and mineralogy. 

White dwarf planetary systems play a crucial role in probing the composition and mineralogy of exoplanetary systems. Exoplanetary bodies from the surviving outer planetary system can be perturbed towards the white dwarf, where they tidally disrupt and form a close-in disc (T$\sim$1000\,K) that subsequently accretes onto the surface of the white dwarf \citep[e.g.][]{debes2002there,jura2003tidally,brouwers2022road}. Emission from this dusty material can be observed in the infrared, and now over one hundred white dwarfs with dust discs are known \citep[e.g.][]{zuckerman1987excess,becklin2005dusty, kilic2006debris, jura2007externally,rebassa2019infrared, wilson2019unbiased, xu2020infrared,lai2021infrared}. These discs are much smaller and hotter compared to the cooler and larger discs around main sequence stars \citep[e.g.][]{Marino2022planetesimal}. Emission from circumstellar gas is also observed for 22 white dwarfs \citep{gaensicke2006gaseous, gansicke2007sdss, gansicke2008sdss, melis2010echoes, farihi2012trio, melis2012gaseous, brinkworth2012spitzer, debes2012detection, dennihy2020five,melis2020serendipitous,gentile2020white, Bhattacharjee2025ZTF}, and absorption from circumstellar gas observed for a handful more \citep[e.g.][]{debes2012detection,xu2016evidence}. The circumstellar environments are notoriously variable showing changes in the dust emission levels \citep{xu2014drop,swan2019most,swan2020dust,swan2021collisions,Guidry2024using,Rogers2025simultaneous}, variations in the equivalent width and morphology of gas emission lines \citep{wilson2014variable,wilson2015composition,manser2015doppler, manser2016another, dennihy2018rapid,manser2019planetesimal,dennihy2020five,gentile2020white,melis2020serendipitous,Rogers2025simultaneous}, and variability in the lightcurves of white dwarfs caused by transiting debris \citep{vanderburg2015disintegrating,Gansicke2016highspeed,vanderbosch2021recurring,Farihi2022relentless,Bhattacharjee2025ZTF, Hermes2025Sporadic}.

The composition of the exoplanetary material in white dwarf planetary systems can be studied in a number of ways: by measuring the bulk elemental composition of the bodies accreted onto the white dwarf photospheres, by measuring the minerals present in their infrared spectra of the circumstellar dust disc, and by constraining the composition of the circumstellar gas. Given that there are more than 1700 polluted white dwarfs with abundance measurements of the exoplanetary material accreted onto the photosphere \citep{Williams2024PEWDD}, this is the more commonly used technique, the majority of which have rocky compositions that match that of Bulk Earth \citep[e.g.][]{zuckerman2007chemical,Dufour2010discovery,swan2019interpretation,xu2019compositions,Trierweiler2023chondritic,doyle2023new,Rogers2024sevenII}. Infrared spectroscopy of the circumstellar discs reveals prominent emission features from minerals present in the disc \citep{jura2009six}, which are most often silicate based. Before the launch of the \textit{JWST}, few white dwarfs with discs had sufficient data for thorough mineralogical analysis of the dust, with G29-38 being the only white dwarf disc with an in-depth mineralogical analysis \citep{reach2009dust}. However, with \textit{JWST}'s sensitivity, the number of white dwarfs with silicate emission features has already doubled \citep{Farihi2025subtle}. Finally, the composition of the circumstellar gas can be studied using photoionisation codes to constrain the composition of the vaporised exoplanetary material \citep{Hartmann2011nonlte,Hartmann2014ton345,gansicke2019accretion,steele2021characterization,Xu2024Cloudy}, and with large spectroscopic surveys on the horizon, the number of white dwarfs with gas discs detected will increase, enabling further studies using this technique. 

Therefore, white dwarf planetary systems provide an opportunity to probe the bulk composition \textit{and} mineralogy of exoplanetary material. The \textit{Spitzer Space Telescope's} InfraRed Spectrograph (IRS) obtained spectra of eight of the brightest white dwarfs with dust discs before the end of \textit{Spitzer's} cryogenic mission, with all spectra showing prominent silicate features around 10\,$\mu$m \citep{reach2005dust,reach2009dust,jura2007infrared,jura2009six}. Since the previous publications, improved data reduction techniques have been implemented and this paper re-examines the infrared spectra to determine the most likely silicate mineralogy for the tidally disrupted planetary material around each of these eight white dwarfs. To compare the bulk elemental composition of the planetary material derived from the white dwarf photospheres with the inferred dust mineralogy, detailed compositional analysis of the white dwarf photosphere are required for each system. Six of the eight white dwarfs with \textit{Spitzer} IRS data have had such an analysis \citep{zuckerman2007chemical,jura2012two,xu2014elemental,gentile2017trace,swan2019interpretation,xu2019compositions}, however, \gd{} and \wdone{} lack sufficient spectroscopic data for the analysis. Therefore, this paper also reports new \textit{Hubble Space Telescope} (\textit{HST}) data which is used to measure the photospheric abundances of the planetary material accreting onto \gd{} and \wdone{}. This then enables for all systems a direct comparison between the silicate mineralogy from the circumstellar dust and the bulk elemental composition (specifically the Mg/Si ratio) from the white dwarf's photosphere. Section\,\ref{Obs} presents the \textit{Spitzer} and \textit{HST} data used in the paper, and discusses the methods used to derive the abundances of the planetary bodies accreted onto \gd{} and \wdone{}. Section\,\ref{Results} reports the analysis of the abundances accreting on to \gd{} and \wdone{}, and compares the bulk composition of the material accreting onto the full sample of eight white dwarfs with the inferred dust mineralogy. Sections\,\ref{Discussion} and \ref{Conclusions} discuss the limitation, implications, and conclusions of these results. 

\section{Observations and Data Analysis} \label{Obs}

\begin{table}
	\centering
	\caption{Properties, abundances, sinking timescales, accretion rates ($\dot{M}_{\mathrm{X}}$), and accreted masses (${M}_{\mathrm{X}}$) for \gd{} and \wdone. The \Teff{} and \logg{} are from \citet{Gianninas2011spectroscopic}, the white dwarf mass ($M_{\textrm{WD}}$) and log(q), q\,=\,log$_{10}$($M_{\textrm{CVZ}}/M_{\textrm{WD}}$), are calculated using models from \citet{Bedard2020spectral}, and the distance is from \textit{Gaia} DR3. }
	\label{tab:WD_Properties}
	\begin{tabular}{lcc} 
		\hline
		 & \gd & \wdone \\
		\hline
        \textit{Gaia} DR3 & 3251748915515143296 & 3571559292842744960 \\
		Atm. & H & H \\
		\Teff~(K) & 15270\err300 & 12640\err200 \\
        \logg & 8.09\err0.10 & 8.22\err0.10\\
        $M_{\textrm{WD}}$ (M$_{\odot}$) & 0.67 & 0.74 \\
        log(q) & $-$16.55 & $-$15.15 \\
        \textit{D} (pc) & 71.6 & 71.3 \\
		\hline
        log $n$(C)/$n$(H) & $-$7.90\err0.10$^{\,2}$& $<-7.57$$^{\,2}$ \\
        $\tau_{\mathrm{C}}$ (yrs) & 0.0105 & 0.0354 \\
        $\dot{M}_{\mathrm{C}}$ (g\,s$^{-1}$) & 1.72$\times 10^4$ & $<3.02$$\times10^5$ \\
        $M_{\mathrm{C}}$ (g) & 5.68$\times10^9$ & $<3.37$$\times 10^{11}$ \\
        
        \hline
        log $n$(O)/$n$(H) & $-$5.54\err0.10$^{\,2}$ & $-$5.50\err0.30$^{\,2}$ \\
        $\tau_{\mathrm{O}}$ (yrs) & 0.00636 & 0.0165 \\
        $\dot{M}_{\mathrm{O}}$ (g\,s$^{-1}$) & 8.64$\times 10^6$ & 1.01$\times10^8$ \\
        $M_{\mathrm{O}}$ (g) & 1.73$\times10^{12}$ & 5.27$\times 10^{13}$ \\

        \hline
        log $n$(Mg)/$n$(H) & $-$5.55\err0.20$^{\,1}$ & $-$6.14\err0.20$^{\,1}$ \\
        $\tau_{\mathrm{Mg}}$ (yrs) & 0.0127 & 0.0231 \\
        $\dot{M}_{\mathrm{Mg}}$ (g\,s$^{-1}$) & 6.44$\times 10^6$ & 2.51$\times10^7$ \\
        $M_{\mathrm{Mg}}$ (g) & 2.57$\times10^{12}$ & 1.83$\times 10^{13}$ \\

        \hline
        
        log $n$(Si)/$n$(H)$_{\mathrm{\,Op}}$ & $-$5.69\err0.20$^{\,1}$ &  \\
        log $n$(Si)/$n$(H)$_{\mathrm{\,UV}}$ & $-$5.49\err0.10$^{\,2}$ & $-$5.93\err0.14$^{\,2}$ \\
        log $n$(Si)/$n$(H)$_{\mathrm{\,Av}}$$^*$ & $-$5.58\err0.22 &  \\
        $\tau_{\mathrm{Si}}$ (yrs) & 0.0103 & 0.0249 \\
        $\dot{M}_{\mathrm{Si}}$ (g\,s$^{-1}$) & 8.55$\times 10^6$ & 4.37$\times10^7$ \\
        $M_{\mathrm{Si}}$ (g) & 2.77$\times10^{12}$ & 3.44$\times 10^{13}$ \\
        \hline
        
        log $n$(P)/$n$(H) & $<-7.63$$^{\,2}$ & $<-5.51$$^{\,2}$ \\
        $\tau_{\mathrm{P}}$ (yrs) & 0.00850 & 0.0221 \\
        $\dot{M}_{\mathrm{P}}$ (g\,s$^{-1}$) & $<5.85$$\times10^4$ & $<1.43$$\times10^8$ \\
        $M_{\mathrm{P}}$ (g) & $<1.57$$\times10^{10}$ & $<9.97$$\times 10^{13}$ \\
        \hline
        
        log $n$(S)/$n$(H) & $<-5.88$$^{\,2}$ & $<-3.49$$^{\,2}$  \\
        $\tau_{\mathrm{S}}$ (yrs) & 0.00652 & 0.0190 \\
        $\dot{M}_{\mathrm{S}}$ (g\,s$^{-1}$) & $<2.08$$\times10^6$ & $<1.80$$\times10^{10}$ \\
        $M_{\mathrm{S}}$ (g) & $<4.27$$\times10^{11}$ & $<1.08$$\times 10^{16}$ \\
        
        \hline
        
        log $n$(Ca)/$n$(H) & $-$6.86\err0.20$^{\,1}$ & $-$7.03\err0.20$^{\,1}$ \\
        $\tau_{\mathrm{Ca}}$ (yrs) & 0.00762 & 0.0145 \\
        $\dot{M}_{\mathrm{Ca}}$ (g\,s$^{-1}$) & 8.65$\times 10^5$ & 8.54$\times10^6$ \\
        $M_{\mathrm{Ca}}$ (g) & 2.08$\times10^{11}$ & 3.90$\times 10^{12}$ \\
        \hline
        
        log $n$(Fe)/$n$(H)$_{\mathrm{\,Op}}$ & $-$5.44\err0.20$^{\,1}$ & $<-$5.70$^{\,1}$ \\
        log $n$(Fe)/$n$(H)$_{\mathrm{\,UV}}$ & $-$5.48\err0.10$^{\,2}$ & $<-$3.86$^{\,2}$ \\
        log $n$(Fe)/$n$(H)$_{\mathrm{\,Av}}$$^*$ & $-$5.46\err0.22 &  \\
        $\tau_{\mathrm{Fe}}$ (yrs) & 0.00541 & 0.0131 \\
        $\dot{M}_{\mathrm{Fe}}$ (g\,s$^{-1}$) & 4.26$\times 10^7$ & $<2.82$$\times10^8$ \\
        ${M}_{\mathrm{Fe}}$ (g) & 7.27$\times 10^{12}$ & $<1.16$$\times10^{14}$ \\
        
        \hline
        log $n$(Ni)/$n$(H) & $-$6.89\err0.10$^{\,2}$& $<-6.72$$^{\,2}$ \\
        $\tau_{\mathrm{Ni}}$ (yrs) & 0.00495 & 0.0113 \\
        $\dot{M}_{\mathrm{Ni}}$ (g\,s$^{-1}$) & 1.82$\times 10^6$ & $<3.27$$\times10^7$ \\
        ${M}_{\mathrm{Ni}}$ (g) & 2.84$\times 10^{11}$ & $<1.17$$\times10^{13}$ \\
        
        \hline
        $\dot{M}_{\mathrm{total}}$ (g\,s$^{-1}$) & 6.89$\times 10^7$ & 1.79$\times 10^8$ \\
        ${M}_{\mathrm{total}}$ (g) & 1.49$\times10^{13}$ & 1.09$\times10^{14}$ \\
        \hline
	\end{tabular}
    \vspace{-7mm}
    \begin{flushleft}
    \item $^1$ Abundances from analysis of optical spectra (Op) from \citet{xu2019compositions}.
    \item $^2$ Abundances from analysis of ultraviolet spectra (UV) from this work. 
    \item $^*$ Average Op and UV abundance, this value is used for subsequent analysis. 
    \end{flushleft}
\end{table}

\subsection{Spitzer IRS data} \label{spitzer-irs}

\begin{figure*}
	\includegraphics[width=0.7\textwidth]{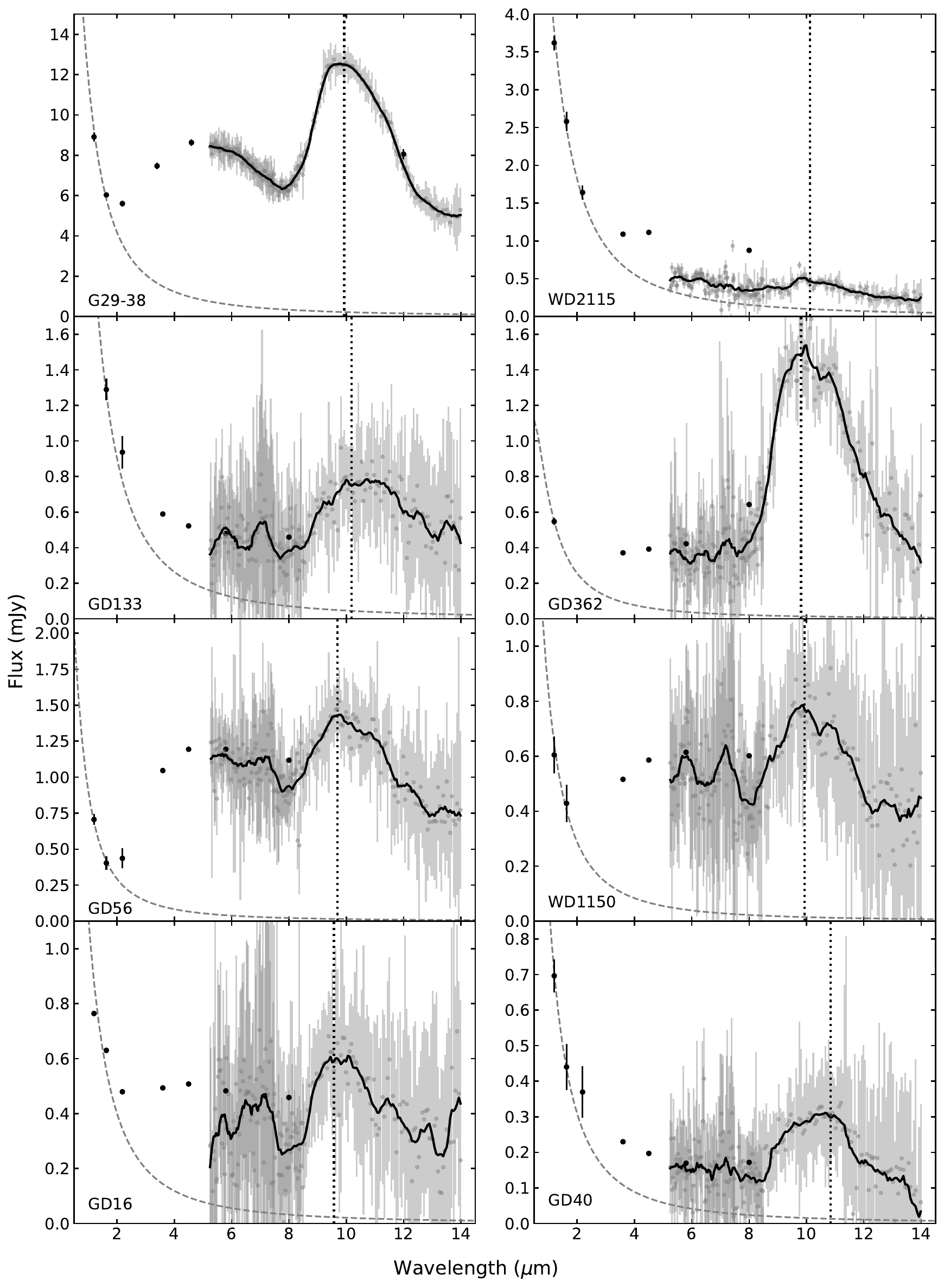}
    \caption{Spectral energy distributions for the eight white dwarfs with \textit{Spitzer} IRS data. Near infrared 2MASS \textit{JHK} and \textit{Spitzer}/\textit{WISE} photometry are shown as black data points. The un-smoothed \textit{Spitzer} IRS data is shown in grey with the data smoothed using a moving point average (box size = 0.75$\mu$m) in black. The dotted vertical lines show where the peak of the silicate emission feature lies after continuum subtracting using a linear model to remove the contribution of the thermal continuum. WD\,2115$-$560 data is unreliable due to observing issues and is shown for completeness.}
    \label{fig:Spitzer_IRS}
\end{figure*}

\textit{Spitzer}/IRS \citep{Werner2004Spitzer,Houck2004IRS} obtained 5-15\,\mum spectra of the eight bright white dwarfs with circumstellar dust that show 10\,\mum silicate emission features during its mission: GD362 (WDJ173134.32+370520.71), GD40 (WDJ030253.10$-$010833.80), G29-38 (WDJ232847.64+051454.24), GD133 (WDJ111912.40+022033.05), \wdone{} (WDJ115315.26$-$153636.51), \gd{} (WDJ041102.17$-$035822.59), GD16 (WDJ014856.80+190227.48), and WD\,2115$-$560 (WDJ211936.53$-$555014.47). These observations were part of the Spitzer programs with the following IDs: 23, 40246, 2313, and 20026. These data were reported and analysed in \citet{reach2005dust,jura2007infrared,jura2009six}. The Combined Atlas of Sources with \textit{Spitzer} IRS Spectra \citep[CASSIS, ][]{Lebouteiller2011CASSIS}\footnote{The Combined Atlas of Sources with Spitzer/IRS Spectra (CASSIS) is a product of the Infrared Science Center at Cornell University, supported by NASA and JPL.} re-processed all stare mode observations to obtain optimal spectral extraction and improved data quality. The CASSIS reduction version LR7 based on the S18.18 \textit{Spitzer} pipeline is used in this study. Figure\,\ref{fig:Spitzer_IRS} shows the spectral energy distributions (SEDs) for the eight white dwarfs including the CASSIS reduction of the \textit{Spitzer}/IRS spectra. The data for WD\,2115$-$560 are unreliable as one of the nods is off the slit, meaning the resulting CASSIS spectra are not trustworthy and are not used in the subsequent analysis.

\subsection{\textit{HST} COS ultraviolet spectra}

\textit{HST} far ultraviolet (FUV) spectroscopic observations of \gd{} and \wdone{} were obtained as part of the programme 17185. The G130M grating was selected with a central wavelength of 1291\,\AA~giving wavelength coverage of 1150--1430\,\AA~(with a 20\,\AA~gap between the two segments). \gd{} was observed for 2 orbits of \textit{HST} on 2023 July 28 and \wdone{} for 6 orbits, split into two non-consecutive visits, on 2023 December 21 and 2023 December 27. The data products from the \textsc{calcos} reduction pipeline were used for subsequent analysis. 

\textit{HST} ultraviolet spectra can be contaminated with airglow from the Earth's atmosphere depending on \textit{HST}'s orbital position. The data show strong geocoronal contributions of Lyman alpha and O\,\textsc{i} emission lines, and so to fit the white dwarf spectra in these regions, the geocoronal emission lines need to be removed. This was done using the STScI COS notebooks\footnote{\url{https://www.stsci.edu/hst/instrumentation/cos/documentation/notebooks}} which separates out the data where the Sun was below the geometric horizon from the point of view of \textit{HST}, this leaves the `night data' which can then be fitted in these regions to obtain the white dwarf abundances for previously obscured lines. To reduce fixed pattern noise, the spectra are taken at different grating offsets, however, given the COS2025 strategy, only 2 offsets, FP-POS 3 and 4, can be used. The exposure times were split between these offsets such that approximately 50 per cent of the data was taken at each offset. The final data was a stack between these offsets and visits weighted by the signal-to-noise ratio (SNR). There are small discrepancies in the resolutions of these offsets, however, this affects the final abundances by less than 0.05 dex \citep{Rogers2023sevenI}. The SNR per resolution element of the stacked data is 20 for \gd{} and 10 for \wdone{} as calculated from the continuum around the C\,\textsc{ii} lines at 1334.530 and 1335.708\,\AA. 

\subsection{Deriving white dwarf abundances} \label{derive_abundances}

\subsubsection{White dwarf parameters:}

The abundances derived from the optical spectra for the material polluting \gd{} and \wdone{} are reported in \citet{xu2019compositions}. To ensure consistency between the analyses, the same parameters are used in this study of the ultraviolet spectra and these were originally derived from spectroscopic observations in \citet{Gianninas2011spectroscopic}. These parameters are reported in Table\,\ref{tab:WD_Properties}.

\subsubsection{Line Fitting:} The equivalent width, equivalent width error, line centre, and radial velocity of each spectral line were measured by fitting a Voigt profile to each line using Markov Chain Monte Carlo to sample through the flux errors. The atomic data bases of Vienna Atomic Line Database (VALD)\footnote{\url{http://vald.astro.uu.se}}, National Institute of Standards and Technology (NIST)\footnote{\url{https://physics.nist.gov/PhysRefData/ASD/lines_form.html}}, and \citet{van2018recent}, as well as published line lists of polluted white dwarfs observed with \textit{HST} \citep[e.g.][]{jura2012two,gansicke2012chemical,Rogers2023sevenI}, were used to identify each spectral line. The radial velocities of the lines were derived using the core of the Voigt profile, and the average photospheric velocity (not including lines from night data or blended lines) is 13.0\,km\,s$^{-1}$ for \gd{} and 21.4\,km\,s$^{-1}$ for \wdone{}. Lines with radial velocities that are significantly discrepant from this are non-photospheric, these absorption lines could result from interstellar medium absorption or absorption from circumstellar material \citep[e.g.][]{debes2012detection,LeBourdais2024revisiting}. The measurements of the lines are reported in Tables\,\ref{tab:gd56-lines} and \ref{tab:WD1150-lines}. For lines and contributions to lines that were non-photospheric in origin, a Voigt profile was filled to the non-photospheric contribution before the white dwarf models were fitted to the photospheric contribution. 

\subsubsection{Abundances:} White dwarf models as described in \citet{dufour2012detailed} were used to derive the abundances of the material polluting the white dwarfs. The spectra were split into 5--15\,\AA \,\,panels around the lines of interest and models were fitted to determine the abundance for each panel. For elements with lines in multiple panels, the average was calculated in linear space to determine the final abundance. Figures \ref{fig:wd1150-models} and \ref{fig:gd56-models} show the final abundance models over-plotted on sections of the \textit{HST} data. The uncertainties are a combination in quadrature of two contributions: a spread error and measurement error. The spread error was calculated from the spread in abundances based on fitting multiple panels for the same element ($\sigma$/$\sqrt N$). If only one line of a particular element is present, the average spread error based on the observations (0.07 dex) was added in quadrature with the equivalent width error. The measurement error was calculated from the equivalent width and equivalent width error of each spectral line (reported in Tables\,\ref{tab:gd56-lines} and \ref{tab:WD1150-lines}) and propagating this through to calculate the associated abundance uncertainty to give the total measurement error. Only lines that were detected at a significance of 3\,$\sigma$ were taken to contribute towards this error, with the exception of oxygen in \wdone{} due to the fact that the lines were measured from the noisier night data. 

\subsubsection{Upper limits:} Upper limits were derived using the same methods as described in \citet{Rogers2023sevenI} where an equivalent width upper limit that would have resulted in a 3\,$\sigma$ spectral line detection was calculated. These equivalent width upper limits are reported in Table \ref{tab:WDs-EW-UL-UV}, and white dwarfs models were used to convert these values into an upper limit on the abundances, and these are reported in Table \ref{tab:WD_Properties}.

\subsubsection{Molecular Hydrogen:} The spectrum of \wdone{} shows molecular hydrogen features at the same radial velocity as the photospheric lines and so originates from its atmosphere. Tens of molecular hydrogen lines are observed, the strongest of which are at wavelengths of: 1345.085, 1356.482, 1366.391, and 1410.636\,\AA. Molecular hydrogen in white dwarfs is commonly observed for objects of this temperature range \citep{Xu2013molecular,Zuckerman2013Hyades}. Caution should be taken when fitting to white dwarfs with molecular hydrogen lines as the wavelengths often overlap with key metal lines. Therefore, for metals in regions dominated by molecular hydrogen, these regions are excluded from the abundance determination and are only used to check for consistency between the data and models.

\subsubsection{Si\,\textsc{iv} lines:} Si\,\textsc{iv} lines have been observed in the ultraviolet spectra of white dwarfs, with some showing a photospheric contribution, an asymmetric tail, or additional blueshifted components \citep[e.g.][]{fortin2020modeling, gansicke2012chemical,Rogers2023sevenI}. The spectrum of \gd{} shows strong absorption lines from Si\,\textsc{iv} that cannot be fitted by the white dwarf models. The 1393.755\,\AA~line is at a radial velocity of $-$20.6\,km\,s$^{-1}$ and the 1402.770\,\AA~line is at a radial velocity of $-$21.3\,km\,s$^{-1}$, which is discrepant from the average photospheric velocity measured to be 13\,km\,s$^{-1}$ (Table\,\ref{tab:gd56-lines}). The difference between these lines and the photospheric velocity is 33.6 and 34.3\,km\,s$^{-1}$ respectively which is close to the gravitational redshift of 34.7\,km\,s$^{-1}$ \citep{Bedard2020spectral}. Therefore, it is likely that the Si\,\textsc{iv} lines are circumstellar in origin. 

\section{Results} \label{Results}

\begin{figure*}
\centering
\subfloat[\gd]{
  \includegraphics[width=0.48\textwidth]{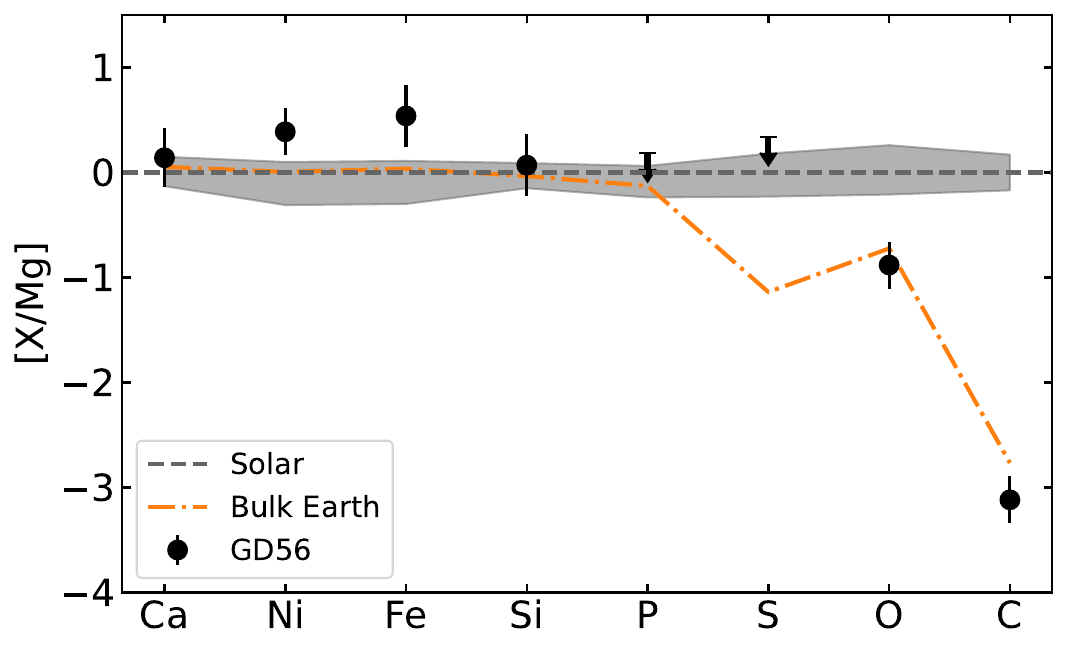}
  \label{fig:gd56-ab}
}
\hspace{2mm}
\subfloat[\wdone]{
  \includegraphics[width=0.48\textwidth]{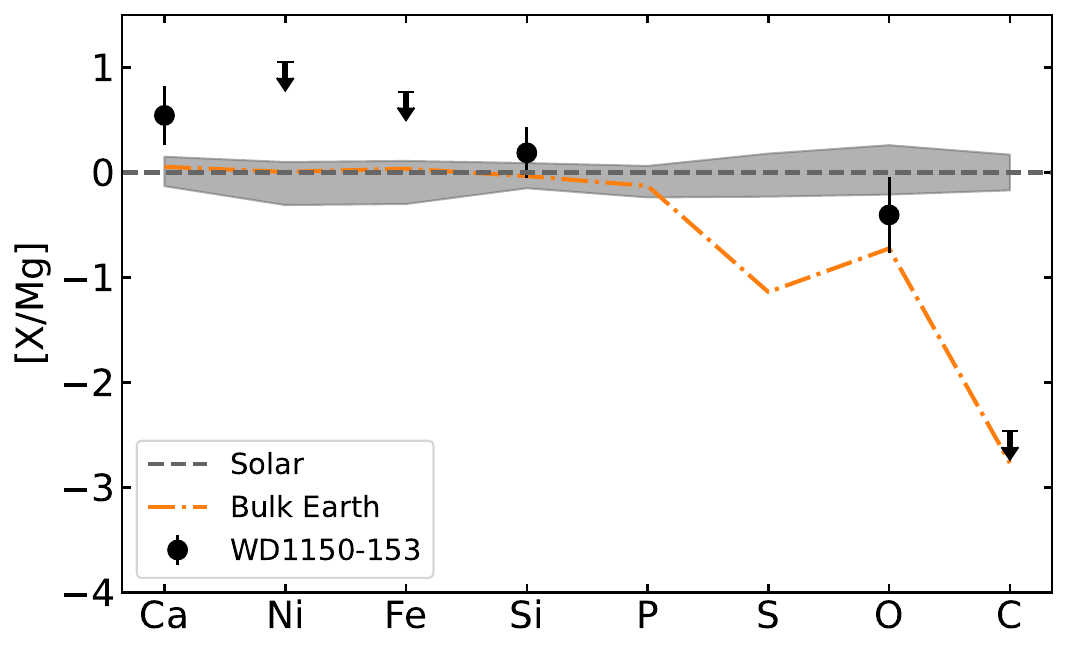}
  \label{fig:wd1150-ab}
}
\caption{The logarithmic number ratio of elemental abundances relative to Mg for the pollutant bodies, normalised to solar (black dashed line) for (a) \gd{} and (b) \wdone. The ratios are plotted assuming the white dwarf is accreting in steady state phase due to the short sinking timescales (Table \ref{tab:WD_Properties}). The grey shaded region around 0 shows the abundance ratios for 95 per cent of nearby main sequence FG stars from the Hypatia catalog \citep{hinkel2014stellar}. Upper limits are plotted as arrows and are measured from the base of the arrow, for \wdone{} the P and S upper limits lie above the range of the y axis.}
\end{figure*}

\subsection{Inferences on the composition of the planetary material accreting onto \gd{} and \wdone}

\subsubsection{Abundance Ratios}

The abundances, sinking timescales\footnote{Sinking timescales are interpolations of grids from \citet{koester2009accretion} using the updated values reported on: \url{http://www1.astrophysik.uni-kiel.de/~koester/astrophysics/astrophysics.html}.}, accretion rates, and accreted masses of each element in the planetary material accreting onto \gd{} and \wdone{} are listed in Table~\ref{tab:WD_Properties}. Ultraviolet spectra reveal the presence of C, O, Si, Fe, and Ni in \gd, while \wdone{} exhibits O and Si. Figures\,\ref{fig:gd56-ab} and \ref{fig:wd1150-ab} show the abundance ratios of the accreted material, normalised to magnesium; magnesium is used as a reference element as it is moderately volatile and has an average condensation temperature in comparison to the other identified elements allowing for easy identification of volatility trends, and it also enables easier identification of differentiation when compared to the siderophilic elements Fe and Ni \citep{harrison2018polluted}. There is a known discrepancy between the pollutant abundances derived from optical and ultraviolet spectra \citep{jura2012two,gansicke2012chemical,xu2019compositions,Rogers2023sevenI}, to account for this, the average abundance in linear space between the previously reported optical data \citep{xu2019compositions} and the ultraviolet measurements are used in the subsequent analysis. The white dwarfs could be accreting in one of three phases, build-up - where accretion has recently begun and the abundance ratios remain largely unaltered by differential settling; steady state - where ongoing accretion requires the abundance ratios to be adjusted to correct for the differential settling of different elements; and declining phase - where accretion has ceased and the element abundances decay exponentially as they sink out of the white dwarf's photosphere. For \gd{} and \wdone{}, the sinking timescales are of the order days to weeks (see Table \ref{tab:WD_Properties}) and they have circumstellar discs that the white dwarfs are accreting from and so it is most likely that these white dwarfs are accreting in steady state as is shown in Figures\,\ref{fig:gd56-ab} and \ref{fig:wd1150-ab}. Steady-state phase is therefore assumed for the subsequent analysis.

The abundance ratios of Ca/Mg and Si/Mg in \gd{} closely match those of the Sun, nearby FG main sequence stars \citep{hinkel2014stellar}, and Bulk Earth. The more volatile elements (O and C) show a depletion relative to solar values, with ratios slightly below those of Bulk Earth. There is a slight enhancement of the siderophilic elements, Fe and Ni, in comparison to Mg. For the Fe/Mg ratio, fewer than 0.05 per cent of FG stars from the Hypatia catalog exceed this value \citep{hinkel2014stellar} making it unlikely that this enhancement is inherited, and for the Ni/Mg ratio, fewer than 0.09 per cent of FG stars exceed this value. These enhancements suggest that \gd{} could be accreting planetary material enriched in siderophilic elements, potentially from an iron core-rich fragment. 

\wdone{} is accreting planetary material with an enhanced Ca/Mg ratio compared to a solar value and a Si/Mg ratio that is consistent with a solar value, nearby stars \citep{hinkel2014stellar}, and Bulk Earth, within error. The O/Mg ratio is depleted in comparison to solar and lies just above that of Bulk Earth. The overall abundance pattern for the elements detected suggests that \wdone{} is accreting dry (oxygen depleted), rocky material with a composition consistent with Bulk Earth.

\subsubsection{\textsc{PyllutedWD}}

To further investigate the origin of the accreted material, the Bayesian framework \textsc{PyllutedWD} \citep{buchan2022planets}\footnote{\url{https://github.com/andrewmbuchan4/PyllutedWD\textunderscore Public}} was used to find the most likely model to explain the planetary material accreting onto these white dwarfs considering sinking effects in the white dwarf photosphere and physical processes that govern the formation and evolution of the planetary material. \textsc{PyllutedWD} starts with a range of initial compositions using nearby stars \citep{brewer2016c} and identifies evidence for geological processes, with the most significant being incomplete condensation and core-mantle differentiation (with subsequent fragmentation). It achieves this by constructing multiple models, each incorporating different geological processes, and applying Bayesian model comparison. When evidence for a geological process is found, \textsc{PyllutedWD} quantifies its statistical significance and constrains the relevant model parameters. 

Analysis with \textsc{PyllutedWD} indicates that \gd{} is most likely in steady state phase and the model that provides the best statistical support given the data (defined as having the highest Bayesian evidence) requires both differentiation and incomplete condensation. \gd{} is therefore accreting planetary material that is dry and core-rich, with incomplete condensation required at a 6.5\,$\sigma$ significance level and differentiation required at a 3.4\,$\sigma$ level. The incomplete condensation model places the temperature at which planetesimal formation occurred (or a temperature it experienced during its lifetime) to be 800$^{+610}_{-410}$\,K. The inferred core mass fraction is 0.59$^{+0.08}_{-0.09}$ (equivalent to a molar core fraction of 0.37$^{+0.08}_{-0.08}$ using the best fitting core composition). These results suggest that \gd{} is accreting material from a differentiated planetary fragment, possibly an iron core-rich remnant.

For \wdone, analysis with \textsc{PyllutedWD} indicates that it is most likely in steady state phase and the best model requires incomplete condensation to a 2.2\,$\sigma$ significance level. \wdone{} is likely accreting material that is dry and \textsc{PyllutedWD} results constrain the temperature at which planetesimal formation occurred to be 550$^{+790}_{-410}$\,K. The associated formation location is 0.63$^{+2.21}_{-0.40}$\,au from its host star \citep[assuming an A0 progenitor star with mass 2.34\,M$_{\odot}$ as these are the likely progenitors,][]{harrison2021bayesian}. This is consistent with Earth's formation location (1\,au) within 1\,$\sigma$, suggesting that \wdone{} is accreting planetary material with a Bulk Earth-like composition. 

\subsection{Dominant silicate mineralogy}

\begin{figure}
	\includegraphics[width=\columnwidth]{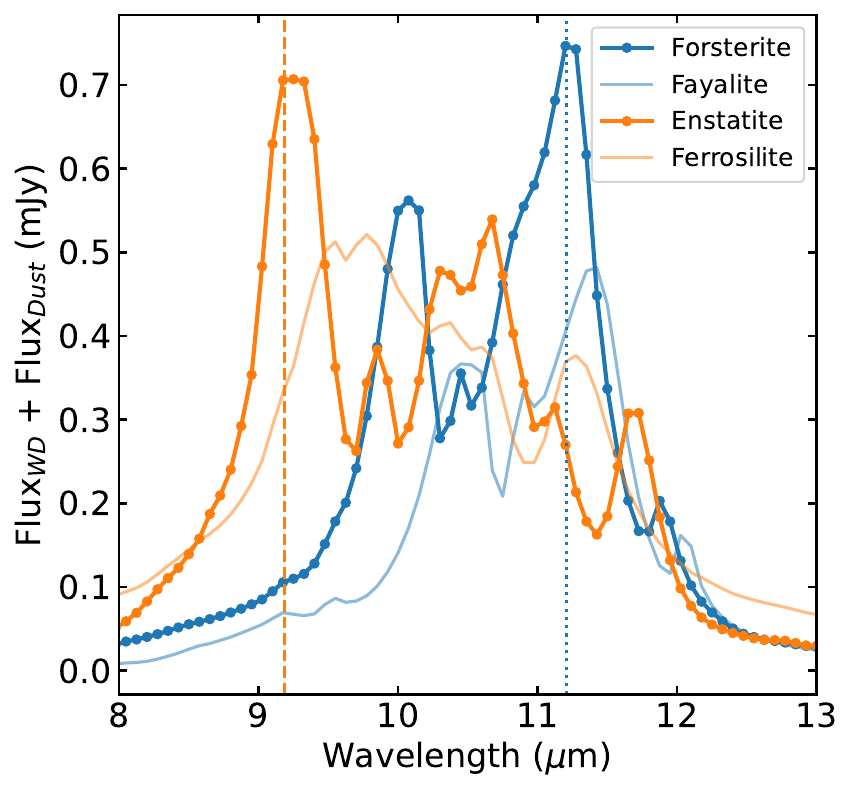}
    \caption{Model spectra for a white dwarf with circumstellar dust with different crystalline silicate minerals sampled at \textit{Spitzer}/IRS resolution. These models assume a DA white dwarf at 100\,pc with an effective temperature 15000\,K and \logg~of 8.0, and optically thin dust located at a single radius such that the dust reaches a uniform effective temperature of 1000\,K, mass of 1$\times 10^{18}$g and different $\kappa_{\mathrm{abs}}$, mass absorption coefficients. The two end members of olivine are shown in blue with Mg rich forsterite: Mg$_2$SiO$_4$ and Fe rich fayalite: Fe$_2$SiO$_4$ using $\kappa_{\mathrm{abs}}$ from \citet{koike2003compositional}, and the two end members of pyroxene are shown in orange with Mg rich enstatite: MgSiO$_3$ and Fe rich ferrosilite: FeSiO$_3$ using $\kappa_{\mathrm{abs}}$ from \citet{chihara2002compositional}. The dashed orange line shows the peak wavelengths for enstatite, peaking around 9.3\micron~and the dotted blue line shows the peak wavelength for forsterite peaking around 11.2\micron. Whether the dust is dominated by olivine or pyroxene changes the peak of the silicate emission feature. Increasing the Fe content of the minerals also move the dominant peak redwards. It should be noted that this assumes the dust is 100 per cent crystalline, whereas, a significant fraction of the silicate emission features are expected to be from amorphous grains.}
    \label{fig:ol-py}
\end{figure}

\begin{figure}
	\includegraphics[width=\columnwidth]{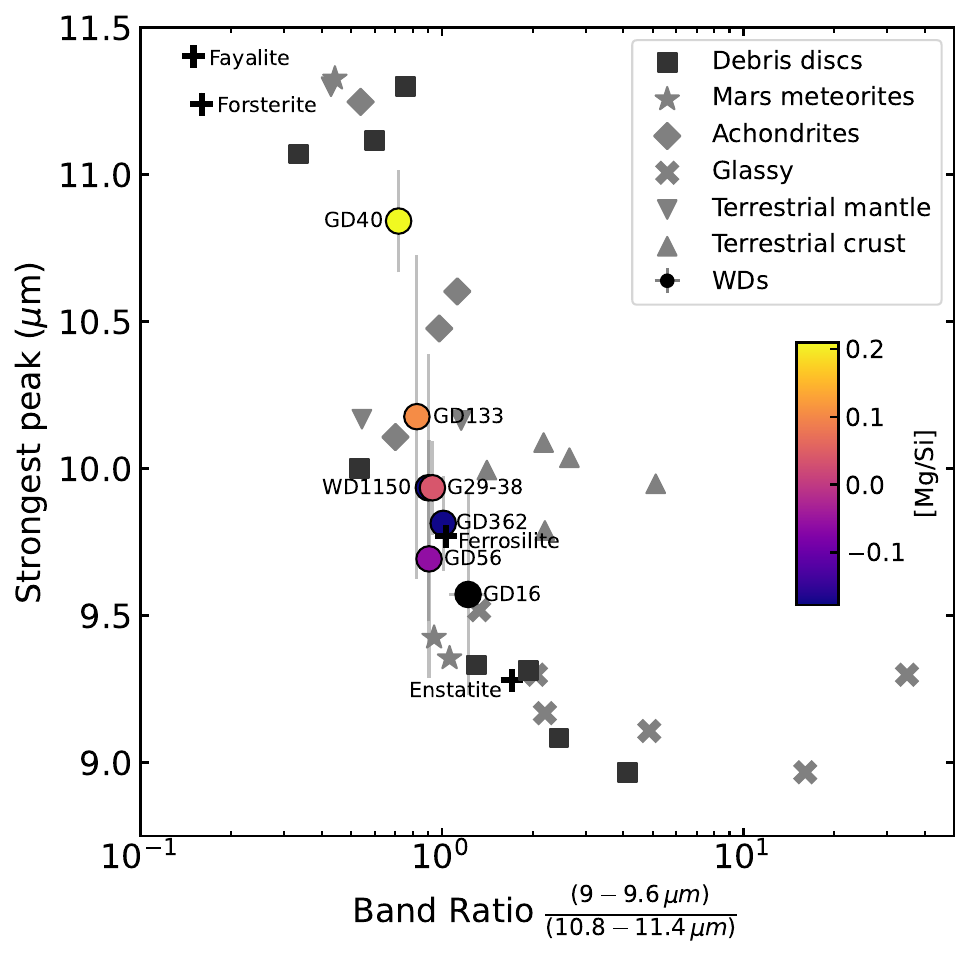}
    \caption{Following the analysis from \citet{Morlok2014dust} showing the band ratios versus peak strength of the silicate emission features for the white dwarfs in the sample compared to extreme debris discs and solar system objects. The band ratio is determined as the integral of the pyroxene dominated band (9-9.6) compared to the integral of the olivine dominated band (10.8-11.4). Crystalline enstatite (Mg-rich pyroxene), crystalline ferrosilite (Fe-rich pyroxene), crystalline forsterite (Mg-rich olivine) and crystalline fayalite (Fe-rich olivine) are shown as `$+$' data points and are labelled for clarity. The colour scale shows the [Mg/Si] ratio in the WD photosphere assuming steady state phase (GD16 appears in black as it is a lower limit). GD40 has the highest [Mg/Si] ratio of the white dwarfs in the sample and is more olivine dominated agreeing with the correlation that the higher the [Mg/Si] ratio, the more olivine dominated the dust should be. This figure uses digitised data from \citet{Morlok2014dust}.}
    \label{fig:morlok-figure}
\end{figure}

Figure\,\ref{fig:Spitzer_IRS} shows the infrared spectra of the eight white dwarfs with \textit{Spitzer} IRS data. As previously highlighted \citep{reach2005dust,reach2009dust,jura2007infrared,jura2009six}, all of these white dwarfs show prominent silicate emission features indicative of optically thin silicate dust. The shape of these features provides insights into the properties of the dust, with the feature width correlating with grain sizes and the peak positions being correlated with the mineral composition \citep[e.g.][]{Ballering2022geometry}.

To interpret these spectra, Fig.\,\ref{fig:ol-py} shows theoretical models of a white dwarf with dust composed of different crystalline silicates. These models assume a DA white dwarf at 100\,pc with an effective temperature of 15000\,K and \logg~of 8.0, along with optically thin micron sized dust with a total mass of $1 \times 10^{18}$\,g located at a single radius such that the dust emits as a single temperature 1000\,K blackbody. The $\kappa_{\mathrm{abs}}$ (mass absorption coefficient) are from laboratory measurements aiming for sub-micron sized grains \citep{koike2003compositional,chihara2002compositional}. The spectra show emission features for crystalline olivine and pyroxene, highlighting the differences between their magnesium-rich (forsterite and enstatite) and iron-rich (ferrosilite and fayalite) end-members. It should be noted that in reality, the dust will contain a significant contribution of amorphous grains and will have contributions from multiple minerals. The enstatite model peaks at 9.3\,$\mu$m, whereas forsterite peaks closer to 11.2\,$\mu$m. Additionally, increasing the iron content in the silicates systematically shifts these peaks toward longer wavelengths. These models reinforce the well-established principle that the dominant silicate mineral in dust can be inferred from the position and shape of the 10\,\mum emission feature. 

Visually comparing these models with the Spitzer/IRS data of the white dwarf dust discs shown in Fig.\,\ref{fig:Spitzer_IRS}, variations in the amorphous-to-crystalline ratio and olivine-to-pyroxene ratio are evident across the sample, with none showing silicate features dominated solely by olivine or pyroxene, but rather a combination of both minerals. Amorphous silicates produce broad featureless emission, whereas, crystalline silicates produce sharper peaks. The amorphous-to-crystalline ratio provides insight into the thermal history of the dust, with high crystalline fractions indicating past thermal processing likely from differentiation, impact processing, or intense stellar irradiation. Notably, the silicate feature of G29-38 appears relatively smooth with a lack of narrow peaks from crystalline silicates, signifying a significant contribution from amorphous silicates as previously highlighted by \citet{reach2009dust,xu2018infrared}. In contrast, the other white dwarfs exhibit more structured peaks in the emission, suggesting a lower amorphous-to-crystalline ratio.  Figure\,\ref{fig:Spitzer_IRS} marks the location of the dominant peaks of the silicate emission features measured from the emissivity profile, where the feature has been normalised by a linear model fitted to the continuum around the silicate emission feature; the dominant peak values are listed in Table\,\ref{tab:WD_Spitzer}. For all dust discs, the dominant peaks fall between the pyroxene (enstatite) peak of 9.3\,\mum and the olivine (forsterite) peak of 11.2\,\mum indicating all dust discs contain a mixture of both minerals. However, there is a clear distinction between systems where the feature peaks closer to the pyroxene side (e.g., GD16 = 9.6\,\micron{} and GD56 = 9.7\,\micron) and those where it peaks closer to the olivine side (e.g., GD40 = 10.8\,\micron), suggesting differences in their olivine-to-pyroxene ratios.

Figure\,\ref{fig:morlok-figure} reproduces Fig.\,4 from \citet{Morlok2014dust}, which compares the peak wavelength of the silicate emission feature with the band ratio, defined as the integral of the pyroxene-dominated band (9-9.6\,$\mu$m) relative to the olivine-dominated band (10.8-11.4\,$\mu$m). As expected, the strongest peak shifts to longer wavelengths with decreasing band ratio separating out objects with different olivine-to-pyroxene ratios. This plot effectively distinguishes the dominant mineralogy of extreme debris discs, Solar System objects, and white dwarf dust discs. To determine the dominant peak and band ratios for the white dwarf dust discs, the silicate feature was normalised using a linear fit to the continuum on either side of the emission feature (7.5-8 and 13-14\,$\mu$m). To assess the sensitivity of these measurements to the normalisation and smoothing methods, the spread in derived peak position and band ratios were calculated by testing different smoothings and normalisation methods. For the normalisation, the following were tested: a single blackbody fitted to the blue side of the silicate emission feature (5-8.5\,\micron), a double black body fitted to both sides of the feature, and a third degree polynomial fitted to both sides of the feature. For the smoothing, different box sizes were used. The standard deviation of the derived values for peak position and band ratio were used to estimate uncertainties.

The white dwarfs predominantly occupy the middle region of the plot, consistent with a mixture of olivine and pyroxene minerals. GD40 stands out as the most distinct case, exhibiting a composition that is notably olivine-rich. None of the white dwarfs overlap with the terrestrial crust region, this matches findings from white dwarf photospheres \citep{Putirka2021polluted} and implies that white dwarfs are not sampling crustal material. Glasses tend to peak around 9\,\micron~and have high band ratios, as seen in some extreme debris discs, but the white dwarfs presented here (i.e., the \textit{Spitzer} sample) do not occupy this region. This may be indicative of white dwarf dust discs lacking glassy compositions, however, recent \textit{JWST} data have identified three white dwarf with strong 9\,\micron~ peaks likely due to glasses \citep{Farihi2025subtle} implying that the absence of such features in this dataset may be an observational bias.

\begin{table*}
	\centering
	\caption{Properties of the eight white dwarfs with \textit{Spitzer} IRS spectra. The [Mg/Si] ratio (log$ \left( \frac{n \mathrm{Mg}}{n \mathrm{Si}}\right)$) is reported for both build-up (BU) and state-state (SS) phases. Measurements for WD\,2115$-$560 are reported for completeness but are not used in the subsequent analysis due to unreliable \textit{Spitzer} data as discussed in Section \ref{spitzer-irs}.}
	\label{tab:WD_Spitzer}
	\begin{tabular}{lcccccccccr} 
    \hline
    Name & H/He & \Teff~(K) & \logg & Mg & Si & [Mg/Si]$_\mathrm{BU}$ & [Mg/Si]$_\mathrm{SS}$ & Peak (\micron) & Band Ratio & Ref. \\
    \hline
    WD\,2115$-$560 & H & 9600 & 7.97 & $-$6.40\err0.10 & $-$6.20\err0.10 & $-$0.20 & $-$0.24 & 10.1 & 0.80 & (1) \\
    GD\,362 & He & 10540 & 8.24 & $-$5.98\err0.25 & $-$5.84\err0.30 & $-$0.14 & $-$0.18 & 9.8 & 1.01 & (2) \\
    GD16 & He & 11000 & 8.0 & $-$7.40\err0.10 & $<-7.20$  & $>-0.20$ & $>-0.24$ & 9.6 & 1.22 & (3) \\
    G29-38 & H & 11820 & 8.4 & $-$5.77\err0.13 & $-$5.60\err0.17 & $-$0.17 & 0.04 & 9.9 & 0.93 & (4) \\
    GD133 & H & 12600 & 8.10 & $-$6.50\err0.20$^\dagger$ & $-$6.60\err0.13 & 0.10 & 0.10 & 10.2 & 0.83 & (4) \\
    \wdone & H & 12640 & 8.22 & $-$6.14\err0.20 & $-$5.93\err0.14 & $-$0.21 & $-$0.18 & 9.9 & 0.90 & (5,7) \\
    \gd & H & 15270 & 8.09 & $-$5.55\err0.20 & $-$5.58\err0.22$^*$ & 0.03
 & $-$0.06 & 9.7 & 0.91 & (5,7) \\
    GD40 & He & 15300 & 8.0 & $-$6.20\err0.16 & $-$6.44\err0.3 & 0.24 & 0.21 & 10.8 & 0.72 & (6) \\
    \hline
	\end{tabular}
    \begin{flushleft}
    \item $^*$ \gd{} Si abundance taken as the average of the optical and ultraviolet values as reported in Table \ref{tab:WD_Properties}.
    \item $^\dagger$ Tentative detection reported so error assumed to be 0.2\,dex.
    \item References: (1) \citet{swan2019interpretation}, (2) \citet{zuckerman2007chemical}, (3) \citet{gentile2017trace}, (4) \citet{xu2014elemental}, (5) \citet{xu2019compositions}, (6) \citet{jura2012two}, (7) This work.
    \end{flushleft}
\end{table*}

\subsection{Comparison between the photosphere and circumstellar dust disc}

A simple yet powerful model correlates bulk Mg/Si to silicate mineralogy by assuming that the elemental Mg/Si number ratio of a rock correlates to whether its silicate mineralogy is olivine or pyroxene dominated. This assumes that the inner regions of the planet forming environment are sufficiently hot that all the material is heated and when forming dust in this environment, all of the initial Mg and Si condenses out leaving little behind in the gas. Higher Mg/Si ratios favour olivine as the dominant mineral due to the two Mg atoms for every one Si atom in olivine (Mg$_2$SiO$_4$) compared to one Mg atom for every one Si atom in pyroxene (MgSiO$_3$). In Earth's upper mantle, where Mg/Si $=1.25$, this model correctly predicts an olivine-dominated composition, consistent with observations from mantle upwellings \citep{McDonough1995composition,Haggerty1995mantle,Palme2003Cosmochemical}.

\begin{figure}
\includegraphics[width=\columnwidth]{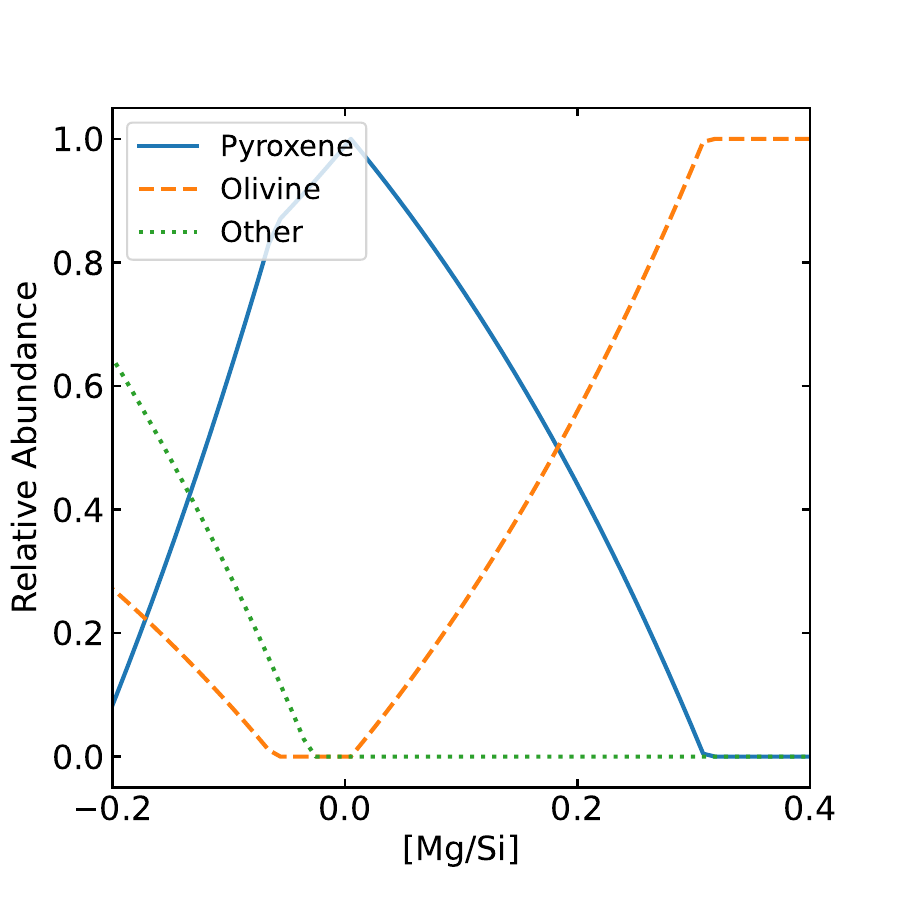}
    \caption{Model output from \textsc{ggchem} showing the relative abundances (compared to all silicate minerals) of olivine (\forsterite{}, orange dashed line) and pyroxene (\enstatite{}, blue solid line) that condense out of a disc with a Bulk-Earth like compositions of Mg, Fe, and O and changing values of Si such that the [Mg/Si] ratio varies from -0.2 to 0.4. The overall behaviour shows that as the [Mg/Si] ratio increases, the relative abundance of olivine increases.}
    \label{fig:ggchem}
\end{figure}

To investigate further how the dominate silicate mineralogy varies as a function of Mg/Si abundance, the equilibrium mineralogy of dust condensing in the mid-plane of a protoplanetary disc was determined for a range of initial compositions using \textsc{ggchem} \citep{Woitke2018ggchem}. \textsc{ggchem} is a chemical equilibrium code that models gas-phase and condensation chemistry. An initial gas-phase composition was inputted and assuming a pressure of 10$^{-5}$\,bar, typical of solids forming in a protoplanetary disc, \textsc{ggchem} minimises the Gibbs free energy to calculate the most stable molecular and mineral species as a function of temperature. Varying Mg/Si ratios were inputted as the initial gas-phase composition to study how the types and proportions of minerals that would condense out from the protoplanetary disc changed as a function of Mg/Si. The Mg/Si ratio was varied by adjusting the initial Si abundances while keeping Mg, Fe, and O fixed at Bulk Earth-like values. Figure\,\ref{fig:ggchem} shows how the dominant silicate mineralogy shifts with increasing Mg/Si for a fixed temperature, here set at 500\,K. For the most Mg-rich cases, the dominant silicate mineralogy is forsterite (olivine, Mg$_2$SiO$_4$) as expected given its higher Mg requirement to form this mineral. As the Mg/Si ratio decreases, the dominate silicate mineralogy transitions to enstatite (pyroxene, MgSiO$_3$) which peaks at a [Mg/Si] ratio of 0 (nMg/nSi = 1). Below 0, mineral formation becomes more complex with the condensation of SiO and SiO$_2$, with the relative amounts depending strongly on the availability of oxygen. These results confirm that, under equilibrium conditions, higher Mg/Si ratios correspond to more olivine-rich mineralogies, in agreement with the simple model outlined above.

\begin{figure*}
\centering
\subfloat[]{
  \includegraphics[width=0.48\textwidth]{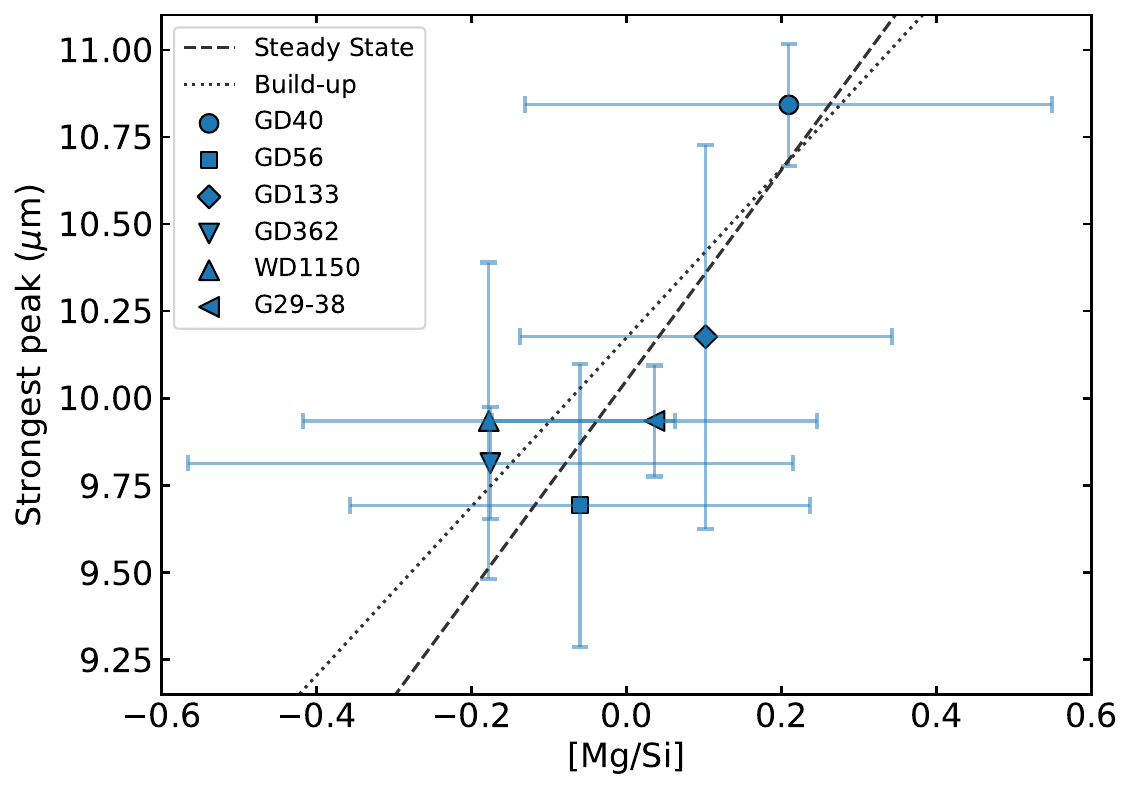}
  \label{fig:Mg_Si_Peak}
}
\hspace{2mm}
\subfloat[]{
  \includegraphics[width=0.47\textwidth]{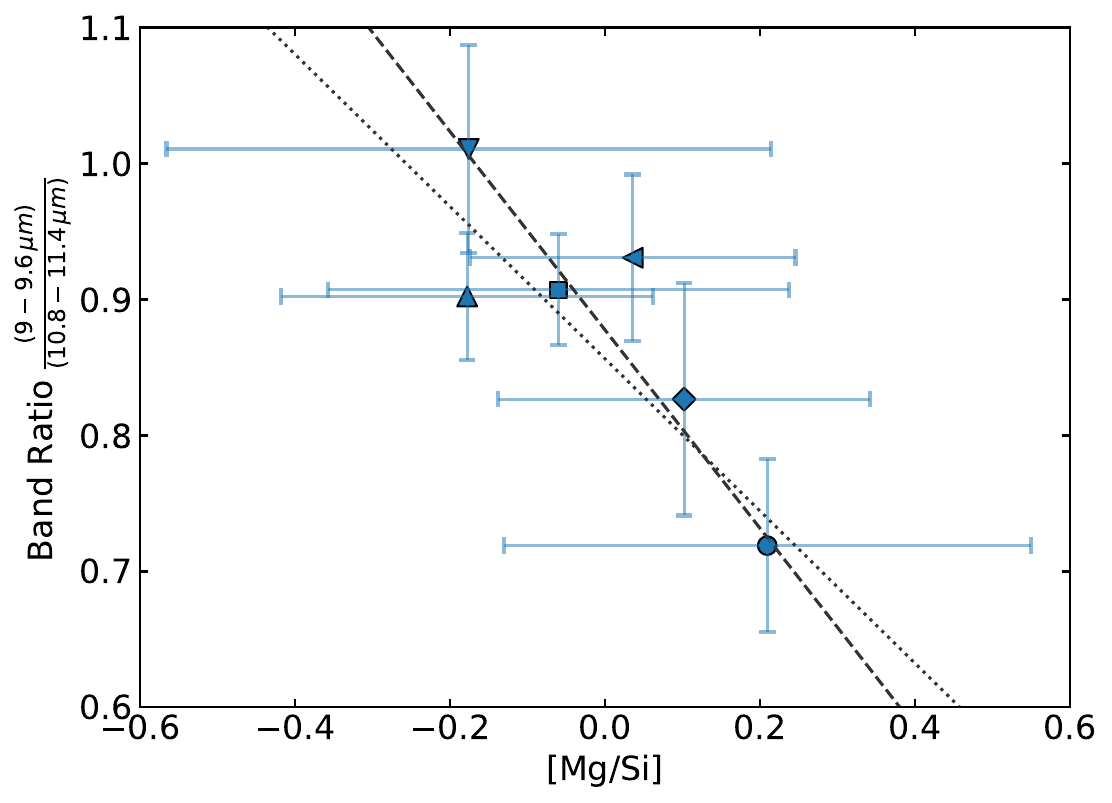}
  \label{fig:Mg_Si_Band}
}
\caption{These plots show (a) the location of the strongest silicate peak and (b) the band ratios (integral of the pyroxene dominated band, 9-9.6\,$\mu$m, compared to the integral of olivine dominated band, 10.8-11.4\,$\mu$m) from Fig.\,\ref{fig:morlok-figure} as a function of the [Mg/Si] steady-state abundance ratio as measured from the photosphere of the white dwarfs. GD40 and GD56 have multiple measurements for their Si abundances and so the averages are used. The dashed line shows the best fitting line derived from orthogonal distance regression assuming the steady-state abundances and the dotted line shows how this best fitting line changes if the (not plotted) build-up phase abundances were instead used. }
\end{figure*}

Polluted white dwarfs provide a unique opportunity to test this relationship, as their photospheres reveal the bulk elemental composition of the accreted planetary material, while their circumstellar dust discs reveal the olivine-to-pyroxene ratio. The colour bar in Fig.\,\ref{fig:morlok-figure} shows the [Mg/Si] ratio of the accreted material for each white dwarf in the sample, assuming steady state phase. Notably, GD40, which has the highest [Mg/Si] ratio, also exhibits the highest olivine-to-pyroxene ratio, supporting the predicted correlation between Mg/Si abundance and dominant silicate mineralogy.

Figures\,\ref{fig:Mg_Si_Peak} and \ref{fig:Mg_Si_Band} show this correlation in more detail demonstrating how the photospheric [Mg/Si] ratio relates to the two key features of the silicate dust emission: the location of the strongest peak, and the ratio of the pyroxene-dominated to olivine-dominated bands. A weighted Pearson's R statistic of 0.80 shows a strong positive correlation between the [Mg/Si] ratio and the location of the strongest peak of the silicate emission feature. A weighted Pearson's R statistic of $-0.73$ shows a strong negative correlation between the [Mg/Si] ratio and the band ratios. It should be noted that this assumes the average abundance measurements for GD40 and GD56, as reported in Table\,\ref{tab:WD_Spitzer}. This demonstrates that there is a correlation between the [Mg/Si] ratio of the exoplanetary material accreted by the white dwarfs and the inferred mineralogy from the dust disc, with the position of the peak wavelength providing a stronger correlation. These correlations also hold if the white dwarfs were instead accreting in build-up phase, as shown by the additional dotted lines in Figures\,\ref{fig:Mg_Si_Peak} and \ref{fig:Mg_Si_Band}.

\section{Discussion} \label{Discussion}

This work compares the mineralogical composition of the circumstellar dust discs and the accreted abundances in the photospheres of eight white dwarfs. For two of these objects, \gd{} and \wdone, new ultraviolet spectra is reported, and an analysis of the abundance pattern reveals that they are accreting dry, rocky bodies, with \wdone{} accreting material that closely matches Bulk Earth's composition, while \gd{} is accreting a core-rich rocky body. By combining this with existing data and analyses of the photospheric composition of six other white dwarfs, the bulk elemental composition of the planetary material accreting onto these white dwarfs are compared to the silicate mineralogy inferred from their infrared spectra from \textit{Spitzer} IRS. These results demonstrate a link between the bulk composition of the rocks accreted by each white dwarf and the mineralogy of these rocks matching predictions from equilibrium chemistry models. The limitations to these conclusions are first discussed below followed by the implications of these results. 

\subsection{Limitations of the analysis methods}

\subsubsection{Deriving white dwarf abundances}

Deriving the abundances of the material accreting onto white dwarfs involves a number of limitations. Specifically to the analysis of \gd{} and \wdone, these are DAZ white dwarfs and in the ultraviolet they typically exhibit fewer metal lines than their DBZ counterparts, leading to fewer spectral lines available for determining metal abundances, with further constraints imposed by the quality of the data. This can bias abundance determination and affect the inferences made on the composition of the accreted material. \wdone{} in particular has few strong metal lines, partly due to the low SNR of the spectra and the many H$_2$ lines overlapping with metal lines. Tables\,\ref{tab:gd56-lines} and \ref{tab:WD1150-lines} report only spectral lines detected at a significance of 3\,$\sigma$, except for the oxygen lines in \wdone, which are fitted using night data that is noisier than the full dataset. Although the derived oxygen abundances should be treated with caution, to change the conclusion that \wdone{} is accreting dry, rocky material, the [O/H] abundance would need to increase by $\sim0.5$\,dex, which is inconsistent with the data.

More generally, a number of systematic uncertainties in white dwarf modeling can result in inaccurate abundance determination, affecting their accuracy in predicting abundance patterns for planetary material and the link with mineralogy. Uncertainties in atomic data can affect the predicted strengths of metal lines as a function of temperature and pressure \citep[e.g.][]{vennes2011pressure}. Additionally, it is well known that the derived metal abundances depend heavily on the assumed white dwarf parameters. This study adopts the spectroscopic derived parameters from \citet{Gianninas2011spectroscopic}, however, for both \gd{} and \wdone{} new photometric parameters for effective temperature and \logg~were derived and the abundances measured assuming these new parameters to check for consistency between parameters derived spectroscopically versus photometrically. Within error, the abundance ratios were not affected, and as the abundance  analysis and comparison with Mg/Si rely on abundance ratios, the variations in white dwarf parameters do not affect the key conclusions. This aligns with previous findings that abundance ratios are less sensitive to stellar parameters than absolute abundances \citep[e.g.][]{Rogers2023sevenI}. Another source of uncertainty arises from systematic discrepancies between metal abundances derived from optical and ultraviolet data \citep{jura2012two,gansicke2012chemical,xu2019compositions,Rogers2023sevenI}, the source of the discrepancy remains debated. The abundances in this work are based on both optical and ultraviolet spectra, introducing potential uncertainties in interpretation. \gd{} has a Si abundance derived from both the optical and ultraviolet data with a discrepancy of 0.2\,dex, but this discrepancy is less than 1\,$\sigma$ and does not impact the key results. The conclusions of magnesium rich dust making olivine rich minerals relies heavily on the GD40 abundances, GD40 has abundances determined for both the optical and ultraviolet giving a [Mg/Si] ratio of 0.52 and 0.23\,dex respectively \citep{jura2012two}. So although the abundance ratios are discrepant, they both show a magnesium rich planetary composition. Finally, atmospheric modelling assumptions introduce another layer of uncertainty, for example, the models used to derive the abundances assume a 1D radial structure for the white dwarf atmospheres, however, 3D effects, such as convective mixing, can influence the inferred accreted abundances \citep{cunningham2019convective}. Although all these limitations will affect the abundances derived, none would result in the key conclusions changing. 

\subsubsection{Deriving dust mineralogy from infrared spectra}

Deriving dust mineralogy from infrared spectra of white dwarf dust discs is challenging. From an observational perspective, the \textit{Spitzer} data is noisy, low resolution, and restricted to wavelengths shorter than 15\,\micron. This makes it more difficult to derive an accurate emissivity profile as the featureless continuum is difficult to measure and this consequently affects the inferred mineralogy. Higher SNR and resolution \textit{JWST} data will provide tighter constraints on the olivine-to-pyroxene ratio. From a modelling perspective, the dominant mineralogy inferred for these dust discs is based on relatively simple spectral analysis methods, which approximate the dominant silicate features but do not fully account for the complexities of dust composition, grain size distributions, or temperature-dependent optical properties \citep{Ballering2022geometry}. In reality, most dust discs are expected to contain a significant fraction of amorphous silicates. Amorphous pyroxene and olivine have different peaks of 9.3 and 9.7\,\micron{} respectively \citep{Dorschner1995silicate,Tsuchikawa2022Spitzer} and so both contribute more flux to the crystalline pyroxene-dominated band (9-9.6\,\micron) compared to the crystalline olivine-dominated band (10.8-11.4\,\micron). Omitting an amorphous silicate component in the modelling of the silicate emission feature therefore systematically biases the inferred olivine-to-pyroxene ratio towards a more crystalline pyroxene composition. In the context of Fig.\,\ref{fig:morlok-figure}, excluding an amorphous component shifts all white dwarf data points towards the lower right corner of the plot, making the discs appear to be more pyroxene-rich. However, this effect would apply systematically across all systems, introducing an offset rather than altering the relative distribution. Therefore, the overall trends and comparative differences between the systems remains robust. Furthermore, including an amorphous component for GD40 would shift it to appearing even more olivine rich, strengthening the conclusion that its dust composition is intrinsically olivine-rich. Future work incorporating more advanced disc models, including radiative transfer and full mineralogical fitting, will be required to determine precise olivine-to-pyroxene and amorphous-to-crystalline ratios. This will enable an improved correlation between the silicate mineralogy and bulk elemental compositions of exoplanetary bodies.

\subsubsection{Assumptions of equilibrium chemistry}

The simple model linking Mg/Si abundance and dominant silicate mineralogy is strongly dependent on the physical and chemical conditions during mineral formation. Therefore, it is assumed that the minerals form under chemical equilibrium and that these equilibrium predictions remain representative of the mineralogy even after millions of years of evolution to the white dwarf phase. Subsequent processing (e.g. thermal) could, however, alter this mineralogy. Mg and Si exhibit similar condensation behaviors and therefore any processing that may occur during this evolution should affect both elements in similar ways. 

Chemically, the Mg/Si ratio is a key factor affecting mantle mineralogy \citep{Hinkel2018star,Spaargaren2020influence,Spaargaren2023plausible}. However, oxygen availability also plays a crucial role in governing the partitioning of iron between metallic and silicate reservoirs, influencing the minerals that form from the disc \citep{Putirka2019composition,Wang2022chemical,Guimond2023mineralogical}. The \textsc{ggchem} models assume a Bulk Earth-like composition for oxygen and iron which may not accurately reflect the true composition of the planet forming disc and may alter the dominant mineralogy predictions from \textsc{ggchem}, impacting the reliability of equilibrium chemistry predictions in this context.

Physically, the \textsc{ggchem} models adopt a protoplanetary disc-like pressure of 10$^{-5}$\,bar, however, the pressure affects the minerals that form. If more extreme pressures and temperatures are experienced either during formation or subsequent evolution then high-pressure forms of minerals such as bridgmanite and ferropericlase become stable and are more likely to form \citep[e.g.][]{Ito1989postspinel,Shim2001post,Guimond2024stars}. This is more relevant for the deep mantle compared to the disrupted bodies accreting by white dwarfs. Observations of polluted white dwarfs find that the majority of accreted objects are asteroid-sized bodies, as indicated by mass estimates of the accreted material and the study of key elements (e.g. Ni, Cr, Si) that show pressure-dependent partitioning behaviors \citep{harrison2021bayesian,buchan2022planets}. Therefore, the higher-pressure mineral phases do not need to be considered for this work and comparing to the low-pressure silicate minerals: olivine and pyroxene, is sufficient. However, for white dwarfs that may be accreting more massive bodies, minerals produced in higher pressure and temperature conditions must be considered. The \textsc{ggchem} models are consistent with findings from \citet{Guimond2024stars} using the modelling package \textsc{pyexoint} \citep{Wang2022chemical} which find that for surface pressures of exoplanets (P$<$15\,Gpa), olivine is the dominant mineral phase at higher Mg/Si ratios.

\subsection{\gd{} and \wdone{} accreting rocky bodies}

\gd{} and \wdone{} are accreting dry, rocky material. Among polluted white dwarfs where the key rock-forming elements (Ca, Mg, Si, Fe, and O) have been detected, the accreted material generally resembles dry, rocky compositions with refractory element abundances matching chondritic meteorites and Bulk Earth \citep[e.g.][]{jura2014extrasolar,harrison2021bayesian,Trierweiler2023chondritic}. Investigations into the oxygen fugacities of these rocks further demonstrate that they most often have an Earth-like geology \citep{doyle2019oxygen, doyle2020extrasolar}. While a handful of white dwarfs have accreted volatile-rich material \citep{farihi2011possible,farihi2013evidence, raddi2015likely, xu2016evidence, xu2017chemical, hoskin2020white, klein2021discovery,Rogers2024sevenII}, these remain the exception rather than the rule. Therefore, \gd{} and \wdone{} add to the growing sample of white dwarfs accreting dry, rocky, material, reinforcing that such compositions are commonplace throughout the galaxy.

The enhanced siderophilic (Fe and Ni) abundances in comparison to lithophilic elements, as well as results from \textsc{PyllutedWD} show \gd{} is likely to be accreting material that is core-rich. This implies that the white dwarf is not accreting the entire body at once in order for a core-rich fragment to be inferred. This adds to just a handful of white dwarfs from the literature that appear to be accreting core-rich material  \citep{melis2011accretion,gansicke2012chemical,hollands2018cool,buchan2022planets,Rogers2024sevenII} and provides further evidence of differentiation and violent collisions in exoplanetary systems. In the Solar System, the decay of short-lived radioactive nuclides (e.g. $^{26}$Al and $^{60}$Fe) causes large scale melting and differentiation in asteroids \citep{urey1955cosmic,elkins2011chondrites,Eatson2024Devolatilization}. Given that most white dwarfs are accreting asteroid sized bodies and that large scale melting from gravitational potential energy likely accounts for less than 0.1 per cent of bodies accreted by white dwarfs, it is likely that  the decay of short-lived radioactive nuclides is the main source of heating for the planetesimals that eventually accrete onto white dwarfs \citep{Jura2013Al26,Curry2022prevalence,Bonsor2023rapid}.

\subsection{Silicate mineralogy and bulk composition of exoplanetary bodies}

Silicate emission features of white dwarf dust discs provide insights into the composition, grain size, and spatial distribution of dust surrounding polluted white dwarfs. This study focused on distinguishing whether the silicate dust is primarily olivine- or pyroxene-dominated for each white dwarf dust disc studied in this \textit{Spitzer} sample. The results indicate that none of the observed white dwarfs exhibit a composition entirely dominated by one mineral phase, suggesting that exoplanetary material contains a mixture of olivine and pyroxene, similar to Earth's mantle. However, there is a clear distinction between systems that are dominated by pyroxene (e.g., GD16 and GD56) and those dominated by olivine (e.g., GD40). These findings reveal that variations in the availability of magnesium and silicon during planet formation can lead to a diversity of minerals across exoplanetary systems.

This study identifies that polluted white dwarfs show a tentative correlation between the [Mg/Si] ratio of an accreted rocky body and its olivine-to-pyroxene ratio. GD40 in particular is accreting planetary material that has the the highest [Mg/Si] ratio of white dwarfs in the sample and has the most olivine-rich dust. This agrees with predictions from \textsc{ggchem} equilibrium chemistry models that planet forming discs that are more magnesium rich are more likely to form olivine minerals. This trend is consistent with observations of Solar System meteorites, where Mg/Si ratios trace the olivine-to-pyroxene fraction in both chondritic and differentiated bodies \citep{Izawa2010Data,Howard2010Data}, albeit over a much smaller Mg/Si ratio range (with nMg/nSi approximately between 0.8 and 1.1). With the observations presented so far this relationship also holds for exoplanetary material, implying that fundamental geochemical processes governing mineral formation are similar across planetary systems. 

\subsection{Connection between the photosphere and circumstellar dust disc} 

The composition of the planetary material in the white dwarfs' photospheres, specifically the Mg/Si ratio, is consistent with the inferred mineralogy of the dust, specifically the olivine-to-pyroxene ratio. This suggests the dust is accreting directly onto the photosphere on sufficiently short timescales. If the timescale for the dust to accrete were longer than the sinking timescale in the photosphere and processes existed to alter the composition (e.g. multiple bodies being accreted, or layer-by-layer accretion), no match would be expected \citep{Johnson2022unusual,Brouwers2023AsynchronousI,Brouwers2023AsynchronousII}. The observed match therefore provides evidence for relatively rapid, and compositionally well-mixed accretion. The absence of significant compositional divergence supports the idea that accretion is occurring on timescales shorter than 1-10$^6$\,years \citep[e.g.][]{girven2012constraints,Veras2020lifetimes}, consistent with steady-state accretion from a circumstellar disc.

\subsection{Mineralogy links to Habitability}

These white dwarf dust disc observations demonstrate the diversity in silicate based minerals that planets form from, with some planets forming from largely pyroxene-based materials whilst in other planetary systems, the planets form from largely olivine dominated minerals. This leads to a diversity in the mineralogy of exoplanets across different stellar systems. This mineralogy controls the prevalent melting behavior of its mantle \citep{Wang2018elemental}, which in turn influences volcanism, mantle convection, and outgassing - essential for sustaining a secondary atmosphere. Pyroxene-rich mantles are more fertile for producing melt in comparison to olivine-rich mantles; this affects the rate of volcanic activity and supply of nutrients to life at the surface \citep{Kelley2010mantle,Lambart2016role}, essential for long term planetary habitability. The ability of a planet's interior to store and cycle water is another key factor impacting habitability. Pyroxene-rich mantles have a higher water storage capacity in comparison to olivine-rich mantles which increases their ability to sustain deep hydrous cycles and sustain water reservoirs \citep[e.g.][]{Keppler2006Thermodynamics}. These mineralogical effects on the mantles lead to further predictions of a diversity in planet density (mass/radius) based on the Mg/Si ratios \citep{Dorn2021water}. It can be speculated that planets with pyroxene dominated mantles may be more likely to be habitable with the volatile storage and cycling necessary for life. 

Determining mantle composition and making inferences about habitability using polluted white dwarfs and their dust discs assumes that they are sampling predominantly mantle material. Although some white dwarfs accrete core-rich material (e.g. \gd), no white dwarf is found to accrete purely core material, and all must have a mantle component. Additionally, \citet{Putirka2021polluted} found that polluted white dwarfs sample the mantle composition over crustal composition. Therefore, polluted white dwarfs and their dust discs are valuable probes of exoplanetary interiors. It is possible to constrain mantle mineralogy directly based on white dwarf photospheric measurements, however, the large abundance errors make this difficult \citep{Trierweiler2023chondritic}. Infrared spectra of the circumstellar dust discs are therefore key to measuring the mineralogy. With \textit{JWST} providing unprecedented access to fainter objects than \textit{Spitzer}, and with \textit{Hubble} observing hundreds of polluted white dwarfs, obtaining statistical samples of exoplanet mineralogy and bulk composition is but a few years away. These findings will feed into exoplanet interior models, linking geochemical properties to observable planetary features, such as atmospheric spectroscopy and surface geology. 

\section{Conclusions} \label{Conclusions}

This work presents new insights into the composition and mineralogy of exoplanetary material accreted by white dwarfs. New ultraviolet spectra of \gd{} and \wdone{} are presented showing they are both accreting dry, rocky material. \wdone{} is accreting material with a bulk composition closely matching that of Bulk Earth, while \gd{} is accreting a core-rich planetary body. These findings contribute to the growing evidence that rocky exoplanetary bodies display a diversity in compositions but with some closely matching Solar System bodies. 

From an analysis of the silicate emission features of the eight white dwarfs observed with \textit{Spitzer} IRS, it was found that all dust discs show a combination of olivine and pyroxene silicate based minerals, as is observed in the Earth's mantle, with GD40 standing out as accreting the most olivine-rich dust. Precise olivine-to-pyroxene ratios remain uncertain due to limitations in current data and analysis methods, with more advanced disc modelling required. 

By comparing the Mg/Si ratio measured in the white dwarf photosphere with the olivine-to-pyroxene ratio inferred from the dust, a correlation was found. The dust and photospheric composition matching implies that the accretion from the circumstellar disc to the photosphere occurs over short timescales in a compositionally well-mixed process. GD40, which is accreting planetary material with the highest [Mg/Si] ratio in the sample, is also the most olivine-rich, supporting that Mg enrichment favors the formation of olivine over pyroxene minerals as predicted from \textsc{ggchem} equilibrium chemistry models. This trend is consistent with correlations observed in Solar System meteorites, suggesting that this relationship extends to exoplanetary material. 

Polluted white dwarfs with circumstellar dust discs can provide key insights into bulk composition and mineralogy of exoplanets, offering a window into the diversity of planetary systems across the galaxy. 

\section*{Acknowledgements}

The authors wish to acknowledge the referee for their useful comments that improved the manuscript. LKR acknowledges Beth Klein for their contributions to the \textit{HST} program number 17185. This research is based on observations made with the NASA/ESA Hubble Space Telescope obtained from the Space Telescope Science Institute, which is operated by the Association of Universities for Research in Astronomy, Inc., under NASA contract NAS 5-26555. These observations are associated with program 17185. LKR and SX are Supported by the international Gemini Observatory, a program of NSF NOIRLab, which is managed by the Association of Universities for Research in Astronomy (AURA) under a cooperative agreement with the U.S. National Science Foundation, on behalf of the Gemini partnership of Argentina, Brazil, Canada, Chile, the Republic of Korea, and the United States of America. KYLS and NPB acknowledge support from NASA XRP 80NSSC22K0234 and 80NSSC24K1486. NPB acknowledges support from NASA grant JWST-GO-03271. 

\section*{Data Availability}

COS FUV (programme ID: 17185) data available from MAST (\href{https://mast.stsci.edu/search/ui/#/hst}{https://mast.stsci.edu/search/ui/\#/hst}). CASSIS \textit{Spitzer} IRS data are available by querying the targets on the following webpage: \href{https://cassis.sirtf.com/atlas/query.shtml}{https://cassis.sirtf.com/atlas/query.shtml}, the Spitzer program IDs are: 23, 40246, 2313, and 20026.
 



\bibliographystyle{mnras}
\bibliography{Master-Bib} 




\appendix

\section{Upper limits}

Following the method discussed in Section \ref{derive_abundances}, the equivalent width upper limits were derived for the strongest line in the wavelength range of the ultraviolet spectra. These values are listed in Table \ref{tab:WDs-EW-UL-UV}.

\begin{table*}
	\centering
	\footnotesize
	\caption{The upper limit equivalent widths using the Hubble COS data in m\AA~of the material polluting \gd{} and \wdone{}. For \wdone{}, the upper limits were derived from data smoothed with a window size of 5 due to the low SNR.}
	\label{tab:WDs-EW-UL-UV}
	\begin{tabular}{ccccccccc} 
		\hline
		Element & Line (\AA) & {\gd}  & {\wdone}  \\
		\hline	
            Fe & 1144.938 & - & 112 \\
            Ni & 1370.132 & - & 30.6 \\
            C & 1335.708 & - & 29.0 \\
            P & 1153.995 & 30.0 & 72.7 \\
            S & 1259.519 & 31.5 & 60.7 \\
  
		\hline
	\end{tabular}
\end{table*}

\section{Interpretation of the \textit{HST} spectra}

For all the spectral lines in the ultraviolet data, a Voigt profile was fitted to derive the equivalent width and radial velocity of the line. These values are reported along with the abundances derived for these lines in Tables \ref{tab:gd56-lines} and \ref{tab:WD1150-lines} for \gd{} and \wdone{} respectively. 

\begin{figure*}
\centering
\subfloat[Si]{
  \includegraphics[width=0.8\textwidth]{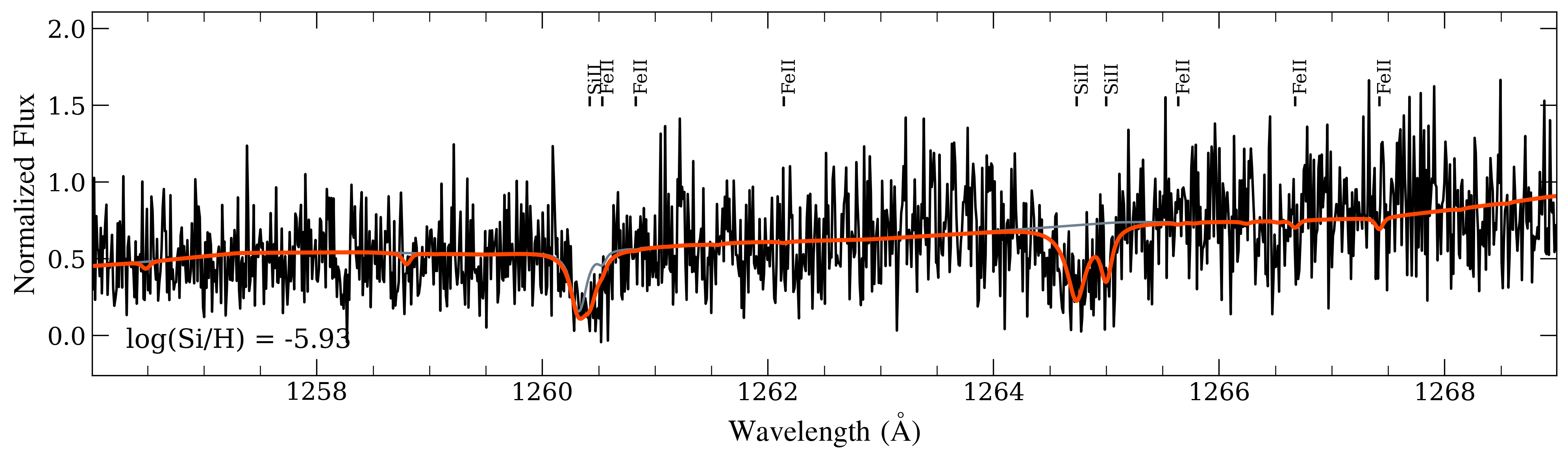}
  \label{fig:wd1150-plots-si}
}
\vspace{2mm}
\subfloat[O]{
  \includegraphics[width=0.8\textwidth]{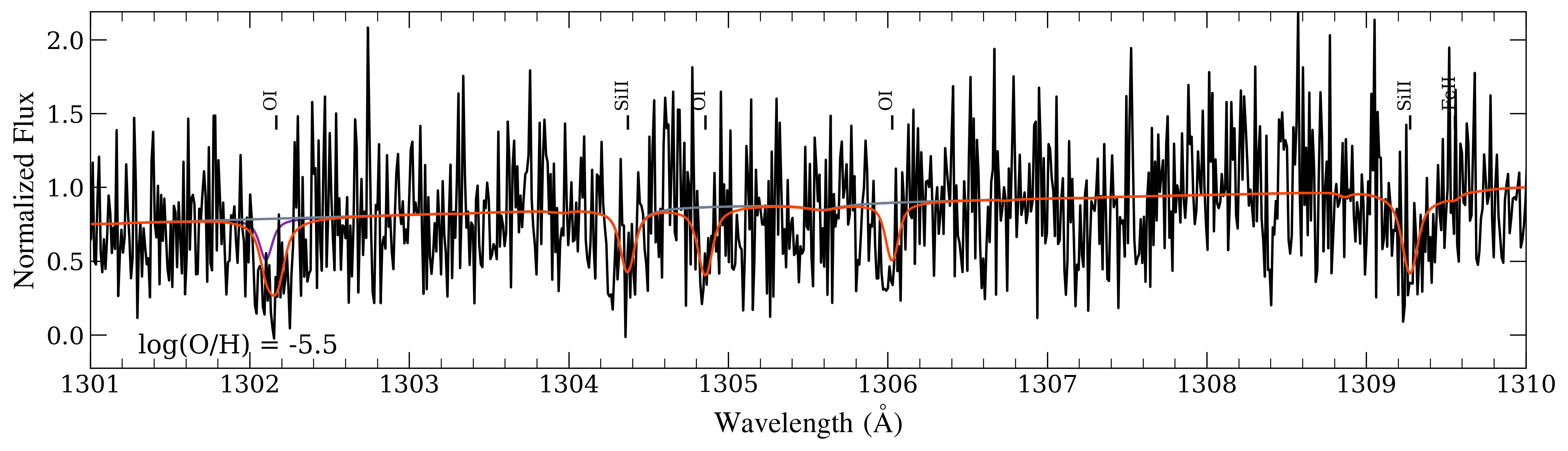}
  \label{fig:wd1150-plots-o}
}
\caption{Model fits to the \wdone{} \textit{HST} spectra, with the data smoothed using a five-point moving average for clarity. Each panel highlights the spectral region around the strongest transitions of a given element: (a) Si lines, and (b) O lines. The grey lines show the model without the specified element included, the purple line shows the non-photospheric absorption features that are modeled using Voigt profiles (for those panels with non-photospheric contributions), and the red line shows a summation of these components and the model fit to the specified element.}\label{fig:wd1150-models}
\end{figure*}

\begin{figure*}
\centering
\subfloat[Fe]{
  \includegraphics[width=0.8\textwidth]{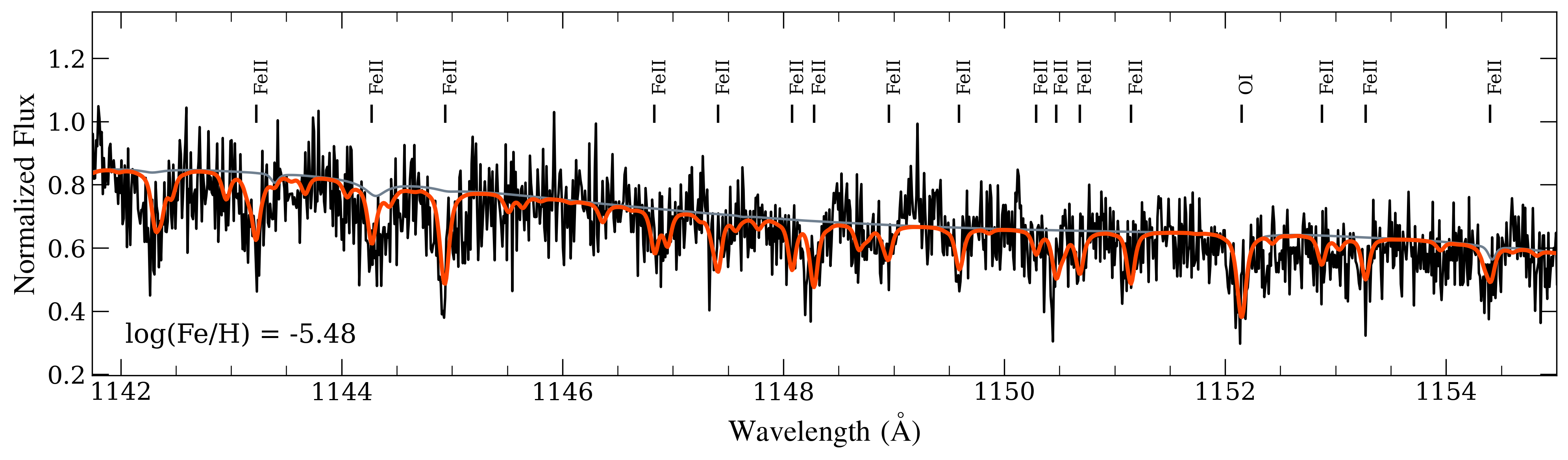}
  \label{fig:gd56-plots-fe}
}
\vspace{2mm}
\subfloat[Si]{
  \includegraphics[width=0.8\textwidth]{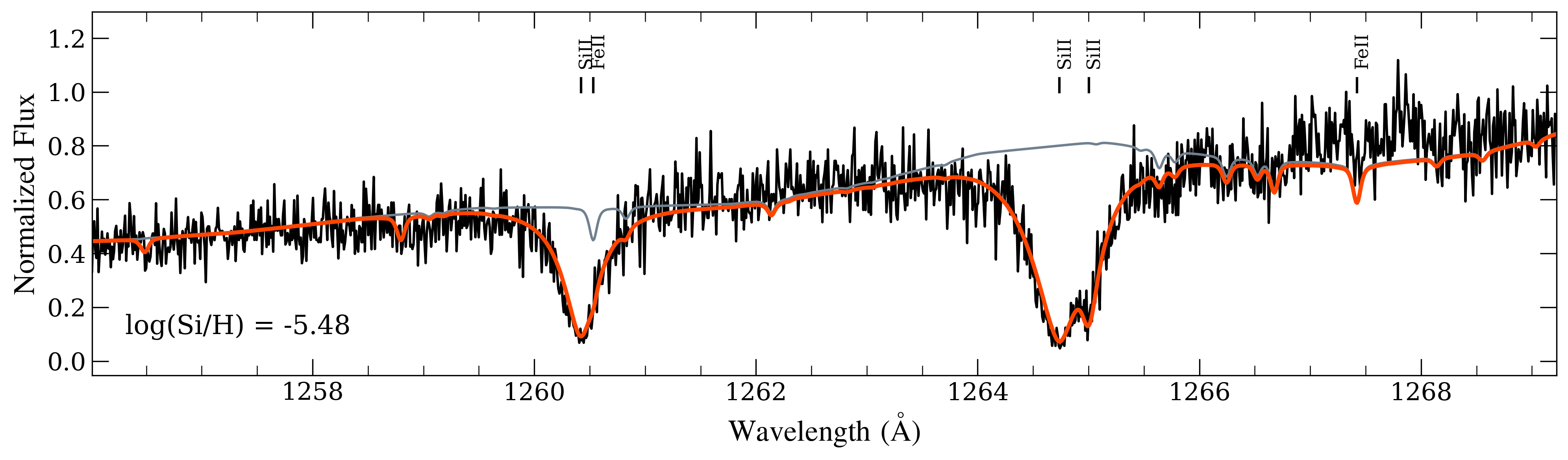}
  \label{fig:gd56-plots-si}
}
\vspace{2mm}
\subfloat[O]{
  \includegraphics[width=0.8\textwidth]{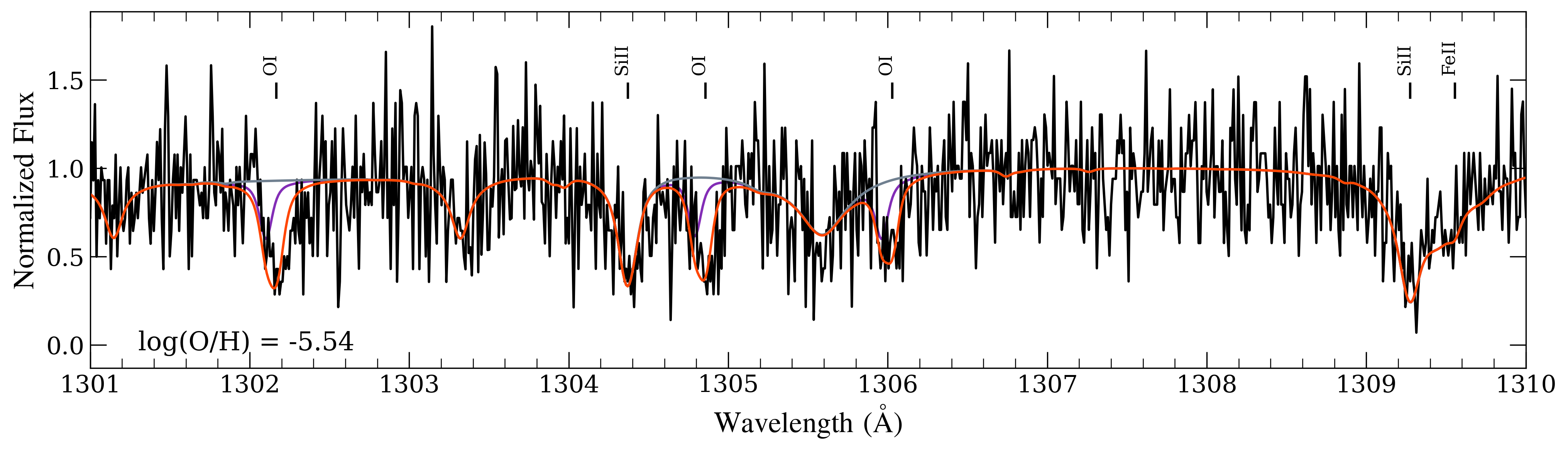}
  \label{fig:gd56-plots-o}
}
\vspace{2mm}
\subfloat[C]{
  \includegraphics[width=0.8\textwidth]{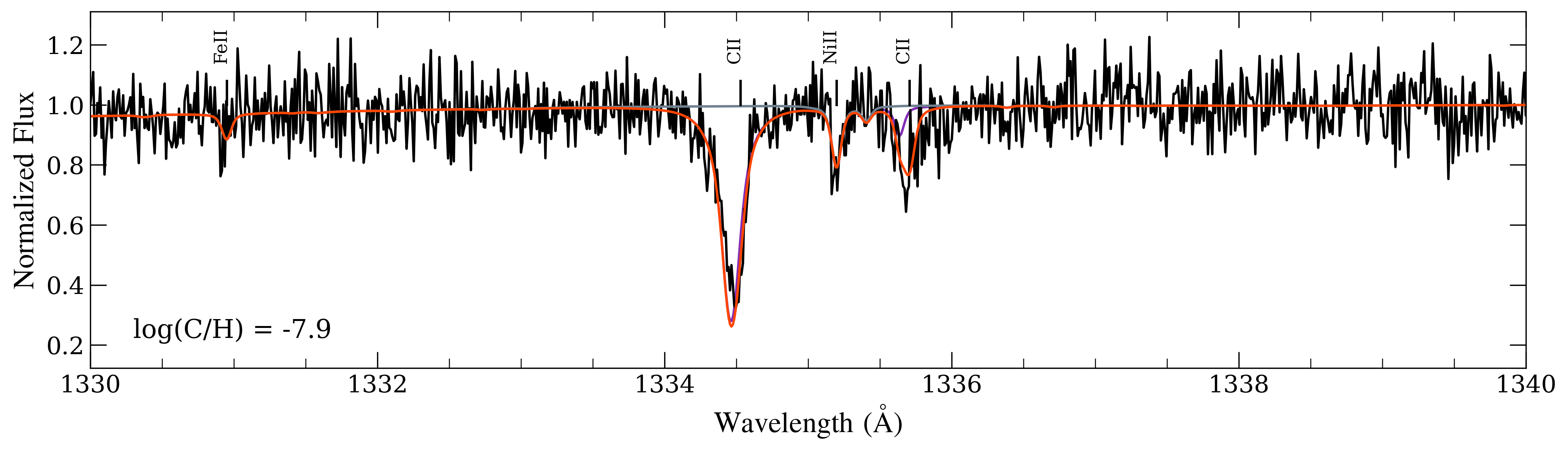}
  \label{fig:gd56-plots-c}
}
\caption{Model fits to the \gd{} \textit{HST} spectra, with the data smoothed using a five-point moving average for clarity. Each panel highlights the spectral region around the strongest transitions of a given element: (a) Fe lines, (b) Si lines, (c) O lines, and (d) C lines (which includes photospheric Ni). The grey lines show the model without the specified element included, the purple line shows the non-photospheric absorption features that are modeled using Voigt profiles (for those panels with non-photospheric contributions), and the red line shows a summation of these components and the model fit to the specified element.}\label{fig:gd56-models}
\end{figure*}

Figures \ref{fig:wd1150-models} and \ref{fig:gd56-models} show the best fitting white dwarf models over-plotted on the \textit{HST} data. 

\begin{table*}
	\centering
	\caption{\textbf{\gd:} The absorption lines in \gd{} from the UV spectrum. The average photospheric velocity (not including lines from night data or blended lines) is 13.0\,km\,s$^{-1}$. If two lines of the same element lie within $\sim$\,10\AA~of one another, the abundances are derived together.}
	\label{tab:gd56-lines}
	\begin{tabular}{lllllllr} 
		\hline
		\hline
		Line & $\lambda _{\textrm{\,vac}}$  (\AA) & log(gf) & E$_{low}$ (eV) & $\lambda _{\textrm{\,obs}}$ (\AA) & RV (km\,s$^{-1}$) & EW (\AA)  & [X/H] \\ \hline \hline
        C\,\textsc{ii} & 1334.530 & $-$0.596 & 0.000 & 1334.580 & 11.2 & 0.1360 (0.0033) & $-$7.90 \\
        C\,\textsc{ii} & 1335.708 & $-$0.341 & 0.008 & 1335.764 & 12.5 & 0.0296 (0.0044) & $-$7.90 \\
        \hline 
        O\,\textsc{i} & 1152.150 & $-$0.268 & 1.967 & 1152.199 & 12.9 & 0.056 (0.013) & $-$5.50 \\
        O\,\textsc{i} & 1302.168 & $-$0.585 & 0.000 & 1302.269 & 23.3 & 0.089 (0.024) & $-$5.58$^{\gamma}$ \\
        \hline
        Si\,\textsc{ii} & 1260.422 & 0.462 & 0.000 & 1260.490 & 16.2 & 0.417 (0.020) & $-$5.43 \\
        Si\,\textsc{ii} & 1264.738$^{\alpha}$ & 0.710 & 0.036 & 1264.870 & 31.2 & 0.687 (0.040) & $-$5.43 \\
        Si\,\textsc{iii} & 1303.323 & $-$0.037 & 6.585 & 1303.420 & 22.3 & 0.54 (0.15) & $-$5.62$^{\gamma}$ \\
        Si\,\textsc{ii} & 1309.276 & $-$0.448 & 0.036 & 1309.380 & 23.9 & 0.172 (0.056) & $-$5.62$^{\gamma}$ \\
        Si\,\textsc{ii} & 1346.884 & $-$0.144 & 5.323 & 1346.948 & 14.3 & 0.0566 (0.0081) & $-$5.44 \\
        Si\,\textsc{ii} & 1348.543 & $-$0.186 & 5.309 & 1348.600 & 12.6 & 0.052 (0.012) & $-$5.44 \\
        Si\,\textsc{ii} & 1350.072 & 0.216 & 5.345 & 1350.129 & 12.7 & 0.0750 (0.0057) & $-$5.44 \\
        Si\,\textsc{ii} & 1352.635 & $-$0.193 & 5.323 & 1352.693 & 12.8 & 0.0383 (0.0075) & $-$5.44 \\
        Si\,\textsc{ii} & 1353.721 & $-$0.158 & 5.345 & 1353.768 & 10.4 & 0.0407 (0.0090) & $-$5.44 \\
        Si\,\textsc{iv} & 1393.755 & 0.030 & 0.000 & 1393.659 & $-$20.6 & 0.5800 (0.0070) & N/A$^{\beta}$ \\
        Si\,\textsc{iv} & 1402.770 & $-$0.280 & 0.000 & 1402.670 & $-$21.3 & 0.4055 (0.0051) & N/A$^{\beta}$ \\
        \hline
        Fe\,\textsc{ii} & 1144.273$^{\alpha}$ & $-$0.94 & 0.083 & 1144.366 & 24.4 & 0.051 (0.011) & $-$5.52 \\
        Fe\,\textsc{ii} & 1144.938 & 0.037 & 0.000 & 1144.980 & 10.9 & 0.0553 (0.0083) & $-$5.52 \\
        Fe\,\textsc{ii} & 1358.937$^{\alpha}$ & $-$0.193 & 3.245 & 1359.036 & 21.9 & 0.0433 (0.0064) & $-$5.52 \\
        Fe\,\textsc{ii} & 1361.373 & $-$0.519 & 1.671 & 1361.449 & 16.6 & 0.0136 (0.0038) & $-$5.49 \\
        Fe\,\textsc{ii} & 1364.578 & $-$0.448 & 3.267 & 1364.618 & 8.7 & 0.0210 (0.0060) & $-$5.49 \\
        Fe\,\textsc{ii} & 1368.094$^{\alpha}$ & $-$0.859 & 1.695 & 1368.141 & 10.2 & 0.0292 (0.0040) & $-$5.49 \\
        Fe\,\textsc{ii} & 1371.022 & $-$0.229 & 2.635 & 1371.066 & 9.7 & 0.0189 (0.0026) & $-$5.49 \\
        Fe\,\textsc{ii} & 1373.718 & $-$0.217 & 3.768 & 1373.770 & 11.3 & 0.0136 (0.0036) & $-$5.49 \\
        Fe\,\textsc{ii} & 1375.172 & $-$0.325 & 2.657 & 1375.227 & 11.9 & 0.0191 (0.0019) & $-$5.49 \\
        Fe\,\textsc{ii} & 1377.987 & $-$0.344 & 3.814 & 1378.044 & 12.4 & 0.0121 (0.0040) & $-$5.55 \\
        Fe\,\textsc{ii} & 1379.470 & $-$0.411 & 2.676 & 1379.526 & 12.1 & 0.0077 (0.0013) & $-$5.55 \\
        Fe\,\textsc{ii} & 1383.580 & $-$0.526 & 2.692 & 1383.640 & 12.8 & 0.0117 (0.0018) & $-$5.55 \\
        Fe\,\textsc{ii} & 1408.478 & $-$0.465 & 2.635 & 1408.571 & 19.9 & 0.0349 (0.0092) & $-$5.36 \\
        Fe\,\textsc{ii} & 1412.842 & $-$1.470 & 0.232 & 1412.892 & 10.7 & 0.0196 (0.0046) & $-$5.36 \\
        \hline
        Ni\,\textsc{ii} & 1317.217 & $-$0.058 & 0.000 & 1317.289 & 16.3 & 0.0269 (0.0048) & $-$6.74 \\
        Ni\,\textsc{ii} & 1335.201 & $-$0.185 & 0.187 & 1335.279 & 17.5 & 0.0182 (0.0038)	& $-$6.93 \\
        Ni\,\textsc{ii} & 1370.132 & $-$0.105 & 0.000 & 1370.199 & 14.6 & 0.0221 (0.0027) & $-$7.05 \\ 
        Ni\,\textsc{ii} & 1381.286 & $-$0.336 & 0.187 & 1381.330 & 9.5 & 0.0166 (0.0023) & $-$7.05 \\
		\hline  \hline
	\end{tabular}
    \begin{flushleft}
    \item $^{\alpha}$\,Blended. 
    \item $^{\beta}$\,Non-photospheric.
    \item $^{\gamma}$\,Extracted using night data.
    \end{flushleft}
\end{table*}

\begin{table*}
	\centering
	\caption{\textbf{\wdone:} The absorption lines in \wdone{} from the UV spectrum. The average photospheric velocity (not including lines from night data) is 21.4\,km\,s$^{-1}$. If two lines of the same element lie within $\sim$\,10\AA~of one another, the abundances are derived together.}
	\label{tab:WD1150-lines}
	\begin{tabular}{lllllllr} 
		\hline
		\hline
		Line & $\lambda _{\textrm{\,vac}}$  (\AA) & log(gf) & E$_{low}$ (eV) & $\lambda _{\textrm{\,obs}}$ (\AA) & RV (km\,s$^{-1}$) & EW (\AA)  & [X/H] \\ \hline \hline
        Si\,\textsc{ii} & 1260.422 & 0.462 & 0.000 & 1260.514  & 21.9 & 0.286 (0.064) & $-$5.82 \\
        Si\,\textsc{ii} & 1264.738 & 0.710 & 0.036 & 1264.864  & 29.8 & 0.295 (0.063) & $-$5.82 \\
        Si\,\textsc{ii} & 1309.276 & $-$0.448 & 0.036 & 1309.373 &  22.2 & 0.120 (0.031) & $-$6.07 \\

        \hline
        O\,\textsc{i} & 1302.168 & $-$0.585 & 0.000 & 1302.253 & 19.6  & 0.22 (0.14) & $-$5.50$^{\alpha}$ \\
        O\,\textsc{i} & 1306.029 & $-$1.285 & 0.028 & 1306.089 & 13.7 & 0.17 (0.12) & $-$5.50$^{\alpha}$ \\

		\hline  \hline
	\end{tabular}
    \begin{flushleft}
    \item $^{\alpha}$\,Extracted using night data. 
    \end{flushleft}
\end{table*}

\bsp	
\label{lastpage}
\end{document}